%% file: sn-article.tex
\documentclass[pdflatex,sn-mathphys]{sn-jnl}%

\usepackage[utf8]{inputenc}
\usepackage{subcaption}
\usepackage{amsmath}
\usepackage[version=4]{mhchem}
\usepackage{siunitx}
\usepackage{longtable,tabularx}
\usepackage{multirow}

\jyear{2022}%

\theoremstyle{thmstyleone}%

\theoremstyle{thmstyletwo}%

\theoremstyle{thmstylethree}%

\input{preamble}

\begin{document}

\title[Computational tool for chemically reacting hypersonic flows]{Development of a high-fidelity computational tool for chemically reacting hypersonic flow simulations}

\author*[1,4]{\fnm{Athanasios~T.} \sur{Margaritis}}\email{a.margaritis@imperial.ac.uk}
\author[2]{\fnm{Cl\'{e}ment} \sur{Scherding}}
\author[3]{\fnm{Olaf} \sur{Marxen}}
\author[4]{\fnm{Peter~J.} \sur{Schmid}}
\author[2,5]{\fnm{Taraneh} \sur{Sayadi}}

\affil[1]{
\orgdiv{Department of Mathematics}, 
\orgname{Imperial College London}, 
\orgaddress{\city{London}, \postcode{SW7 2AZ}, \country{United Kingdom}}}
\affil[2]{
\orgdiv{Institut Jean Le Rond d’Alembert}, 
\orgname{Sorbonne Universit\'{e}, CNRS}, 
\orgaddress{\city{Paris}, \postcode{75005}, \country{France}}}
\affil[3]{
\orgdiv{Department of Mechanical Engineering Sciences}, 
\orgname{University of Surrey}, 
\orgaddress{\city{Guildford}, \postcode{GU2 7XH}, \country{United Kingdom}}}
\affil[4]{
\orgdiv{Department of Mechanical Engineering}, 
\orgname{King Abdullah University of Science and Technology}, 
\orgaddress{\city{Thuwal}, \postcode{23955}, \country{Saudi Arabia}}}
\affil[5]{
\orgdiv{Institute of Combustion Technologies},
\orgname{RWTH-Aachen University},
\orgaddress{\city{Aachen}, \postcode{52062}, \country{Germany}}}

\abstract{
In this paper, we present a methodology to achieve high-fidelity simulations of chemically reacting hypersonic flows and demonstrate our numerical solver's capabilities on a selection of configurations. The numerical tools are developed based on previous in-house codes for high-speed simulations with improvements in both numerical and physical modeling.  Additionally, a modular, open-source library is coupled with the flow solver for modeling real-gas effects in variable atmospheric mixtures. Verification against literature is done for canonical flat-plate boundary layers with various choices for gas modeling, with excellent agreement observed in all cases. The implementation of an artificial-diffusivity shock-capturing numerical scheme is then verified for supersonic shockwave--boundary-layer interaction (SBLI) cases and the improved code's capabilities are demonstrated for the cases of hypersonic SBLI and a sonic jet injection in a hypersonic crossflow, at higher enthalpy levels than those previously investigated. The results show excellent agreement with previous observations in the literature. The work presented in this paper demonstrates the range of applications that can be investigated with this tool, highlights the need for accurate physicochemical modeling, and paves the way for addressing increasingly more complex configurations and flows.
}


\maketitle

\section{Introduction}\label{sec:intro}

Recent advances in the aerospace sector have pushed the limits of vehicle design to far higher speeds and more extreme conditions than ever before. The revival of interest for complex orbital and inter-planetary missions, and the pursuit of commercial hypersonic flight, create the need for accurate and reliable atmospheric (re-)entry and hypersonic cruise vehicle design. Hypersonic boundary layers have posed a significant research and engineering challenge in this context, with fundamental research in hypersonic aerothermodynamics becoming a necessity \citep{BertinCummings_2003,Schmisseur2015,Leyva2017,Theofilis_2022}.

The flow environment encountered near the walls of a vehicle or object traveling at hypersonic speeds is an extremely complex problem, where the interfaces between disciplines become unclear. The interplay between thermodynamics, chemistry, and fluid mechanics creates an interdisciplinary problem that needs to be tackled by accounting for a variety of interconnected phenomena. Even in cases where each of these phenomena is well understood on its own, their combination is not always easy to model, predict, and interpret, as highlighted by a recent article \citep{Leyva2017} discussing these interactions for high-enthalpy high-speed flows; interactions which had already been recognized three decades earlier \cite{Holden1986}.

It is notoriously difficult and expensive to collect reliable experimental and in-flight data for hypersonic flows \cite{Schneider2008}. Due to the cost and difficulty of reliable experimental studies, accurate and reliable numerical simulations to characterize and predict the flow environment become even more indispensable. Typical numerical codes and tools fail to describe the complex and intricate nature of hypersonic flow, either due to unsuitable numerical techniques, or simply the lack of models for the multiphysical aspects of the flow.

The need for accurate and robust numerical simulations of hypersonic boundary layers has led researchers to the development of computational tools that solve the governing equations in the hypersonic flow regime. Initially focusing on the stability and transition of boundary layers, early investigations have tried to extend concepts from subsonic and supersonic flow to the hypersonic regime for transition and instability evolution prediction \cite{Malik1989,Malik1990,Malik1991,Chang1997}. The recent work at the \VKI\ has provided thorough investigations of the effects of chemistry \cite{Zanus2017,MiroMiro2017,MiroMiro2018,Wagnild2018,Beyak2018,Wartemann2019,MiroMiro2019a} and surface features, such as roughness, curvature \citep{Zanus2018,Zanus2019}, and outgassing \citep{Pinna2017,MiroMiro2018a,MiroMiro2019,MiroMiro2019a}, on the stability of boundary layers, using linear and nonlinear \citep{Pinna2017,Zanus2017,MiroMiro2017,MiroMiro2018,MiroMiro2018a,Wagnild2018,Beyak2018,Zanus2019,Wartemann2019,MiroMiro2019,MiroMiro2019a,Zanus2018} stability analysis tools. An important outcome is the conflicting results these authors have sometimes found, which indicate, as they note, the immaturity of our understanding at this stage, concluding that incorrect modeling of transport phenomena could be as inaccurate as neglecting chemical activity altogether. Similar work has been conducted by the group of Candler using \LST\ \cite{Hudson1997, Johnson1998} and \PSE\ \cite{Johnson2005} for reacting flows, extending to shape optimization of hypersonic bodies \citep{Johnson2008,Nichols2019}.

The \TCNEQ\ effects are a significant modeling challenge in hypersonic flow simulations. At such high-energy gas states and given the extremely short time scales of hypersonic flows, there is no guarantee that collisions are frequent enough for energy exchange and chemical processes to reach equilibrium. Therefore, the composition and properties of the gases vary in space and time, following the relevant kinetics and thermodynamics. This creates the need for far more sophisticated gas models for the thermodynamic and transport properties of such gases. The significance of finite-rate phenomena is summarized in a recent comprehensive review \cite{Candler2019}. In the continuum regime, the Navier-Stokes equations still hold for hypersonic reacting gases, with some necessary modifications for the gas properties and the addition of extra equations to model the chemical reactions and track the gas composition \cite{Josyula2015_equations}. 

State-of-the-art tools for \TCNEQ\ models have been developed to simulate the time evolution of the governing equations. A variety of powerful tools based on finite-volume techniques and \MT\ models have been developed based on the work of \citet{Candler1991}, leading to the numerical codes \acrshort{DPLR}, using the \DPLR\ method \citep{Wright1998}, and \acrshort{US3D} \citep{Candler2015, Josyula2015_cfd}. Such codes can be used for extracting the base flow for stability calculations or to simulate directly the nonlinear evolution of the flow \cite{Subbareddy2014,Shrestha2019}.  Such \DNS\ results reveal the underlying mechanisms of transition and serve as benchmarks for lower-fidelity prediction methods. More recent results have extended such simulations further into the turbulent regime \citep{direnzo2020_htr,direnzo2021_htr12,direnzo2021_dns,passiatore2021_prf,sciacovelli2021_shock,passiatore2022}.
Other approaches have used high-order finite-difference \LES\ and \DNS\ solvers for such studies \citep{Marxen2008, Marxen2010a, Ghaffari2010, Groskopf2010, Marxen2010, Marxen2011, Marxen2013, Marxen2014a, Marxen2014}. 
Initial efforts have been made to include real-gas effects and \CNEQ\ modeling in the simulations by these authors.

The objective of this work is to develop a versatile computational tool that is able to accurately model hypersonic boundary layers including a number of phenomena involved in realistic problems, such as finite-rate chemistry effects and shockwaves.
For that, we extend a high-order finite-difference tool \cite{nagarajan2003, nagarajan2004_thesis} to include finite-rate chemistry modeling and shockwaves, building on previous work by \citet{sayadi2013} and \citet{Marxen2014}.
The real-gas and finite-rate chemistry effects are included by coupling our numerical tool with the \acrshort{Mutation++} library \cite{Scoggins2020,Scoggins2017}, an open-source library developed by the \VKI\ that is able to accurately calculate the thermodynamic, transport, and chemical kinetics properties of various high-enthalpy weakly-ionized gases. 
Verification of our computational tools is performed against previous results, while some new results are presented that showcase its capability to handle more complex cases. This novel computational tool is capable of including a multitude of physics that are very rarely dealt with simultaneously by numerical codes.

This paper is organized as follows. The fundamentals of hypersonic fluid dynamics are presented in \cref{sec:theory}, including the governing equations and basic modeling aspects. The numerical techniques employed are described in \cref{sec:numerics}. Code verification and validation, and new results for more complex cases are discussed in \cref{sec:results}. Finally, the main results are emphasized and conclusions are drawn in \cref{sec:conclusions}.

\section{Background theory}\label{sec:theory}

In the following, we describe the general equations governing the flow problems discussed in this work. Our starting point is an overview of the conservation equations, followed by a detailed discussion of the various terms and modeling choices.

\subsection{Governing conservation equations}\label{sub:goveq}

The nondimensional Navier-Stokes equations for fluids that consist of a mixture of species, \speciesset, are presented in \crefrange{eq:globalmass}{eq:totenergy}. \Cref{eq:globalmass} is the continuity equation, describing the global mass conservation in the system. \Cref{eq:speciesmass} corresponds to the set of mass conservation equations for each species, with the net production rate terms, \massprod, appearing on the right-hand side. For non-reacting gas mixtures, where the mixture composition can be considered either constant or a direct function of the thermodynamic state, only the global mass conservation, \cref{eq:globalmass}, is needed; the species mass conservation equations, \cref{eq:speciesmass}, are not necessary and can be omitted in this case. In order to ensure global mass conservation, in the case of a finite-rate reacting mixture with a varying composition, \cref{eq:globalmass} needs to be solved together with \cref{eq:speciesmass} for all but one species. The omitted species is selected based on numerical considerations, commonly avoiding species with the smallest concentrations. Alternatively, all the species mass conservation equations can be solved while \cref{eq:globalmass} is relaxed. For gases in \TNEQ, in which case different energy levels are treated independently, additional energy conservation equations are needed; these cases are not investigated in this work, since our attention is focused on cases where \CNEQ\ effects are dominant.

\begin{alignat}{4}
	&\pd{\density}{\timevar} 
	&&+ \divgp{\density \velocity}
	&&= 0
    \label{eq:globalmass} \\
	\left\{\vphantom{\frac{1}{1}}\right.
	&\pd{\densityspecies}{\timevar} 
	&&+ \divgp{\densityspecies \velocity + \densityspecies \diffvelocity}
	&&= \massprod
	\left.\vphantom{\frac{1}{1}}\right\} \;,\;
	\forall \; \speciess \in \speciesset
	\label{eq:speciesmass} \\
	&\pd{\density \velocity}{\timevar}
	&&+ \divgp{\density \velocity \otimes \velocity}
	&&= - \grad{\pressure} + \divg{\viscstresstensor} 
	\label{eq:momentum} \\
	&\pd{\density \totale}{\timevar}
	&&+ \divgp{\density \totalh \velocity}
	&&= \divgp{\viscstresstensor \cdot \velocity} - \divg{\heatflux} 
	\label{eq:totenergy}
\end{alignat}

The nondimensional quantities are time, \timevar, density, \density, velocity, \velocity, pressure, \pressure, stress tensor, \viscstresstensor, total energy, \totale, total enthalpy, \totalh, and heat flux, \heatflux, along with the partial density, \densityspecies, the net mass production rate, \massprod, and the diffusion velocity, \diffvelocity, for species \speciess. More details regarding the derivation of the equations and the validity of assumptions are provided in \cite{Anderson2019,Gnoffo1989,Josyula2015_equations}.

The governing equations, \crefrange{eq:globalmass}{eq:totenergy}, can be brought into the following compact form, seen in \cref{eq:goveqflux,eq:goveqfluxdesc}. 

\begin{equation}
    \pd{\statevec}{\timevar} 
	=
	- \divg \fluxvecconv
	+ \divg \fluxvecdiff
	+ \sourcevec
	=
	\fluxvec\left(\statevec\right)
	\label{eq:goveqflux}
\end{equation}

\begin{equation}
\small
    \statevec = \left[ 
        \begin{array}{c}
            \density \\ 
            \densityspecies_{_1} \\ 
            \vdots \\ 
            \densityspecies_{_\numspecies} \\ 
            \density \velocity \\ 
            \density \totale
        \end{array} \right]
    ,\,
    \fluxvecconv = \left[ 
        \begin{array}{c}
            \density \velocity \\ 
            \densityspecies_{_1} \velocity \\ 
            \vdots \\ 
            \densityspecies_{_\numspecies} \velocity \\ 
            \density \velocity \otimes \velocity + \pressure\\ 
            \density \totalh \velocity
        \end{array} \right]
    ,\,
    \fluxvecdiff = \left[ 
        \begin{array}{c}
            0 \\ 
            - \densityspecies_{_1} \diffvelocity_{_1} \\
            \vdots \\ 
            - \densityspecies_{_\numspecies} \diffvelocity_{_\numspecies} \\ 
            \viscstresstensor \\ 
            \viscstresstensor \cdot \velocity - \heatflux
        \end{array} \right]
    ,\,
    \sourcevec = \left[ 
        \begin{array}{c}
            0 \\ 
            \massprod_{_1}\\ 
            \vdots \\ 
            \massprod_{_\numspecies} \\ 
            0 \\ 
            0
        \end{array} \right]
    \label{eq:goveqfluxdesc}
\end{equation}

As mentioned above, there are $\numspecies$ conservation equations for the species, so the total problem involves $\numspecies+5$ equations. However, only $\numspecies-1$ of the $\numspecies$ species conservation equations need to be solved. Therefore, the final problem involves $\numspecies+4$ scalar equations.

The nondimensionalization of the equations is done with the reference parameters in \cref{eq:nondimensionalization}, where $\valref{\temperature} = (\freestream{\gammaratio}-1) \freestream{\temperature}$ and the speed of sound is \soundspeed.

\newcommand{\nondimensionalize}[2]{\ensuremath{#1 = \frac{\dimval{#1}}{#2}}}
\begin{equation}
\begin{alignedat}{1}
    &
    \nondimensionalize{\density}{\freestream{\density}}
    ,\;
    \nondimensionalize{\pressure}{\freestream{\density}\freestream{\soundspeed}^2}
    ,\;
    \nondimensionalize{\temperature}{\valref{\temperature}}
    ,\;
    \nondimensionalize{\velocity}{\freestream{\soundspeed}}
    ,\;
    \nondimensionalize{\densityspecies}{\freestream{\density}}
    ,\;
    \vec{V}_\speciess= \frac{\dimval{\vec{V}}_\speciess}{\freestream{\soundspeed}}
    \\
    &
    \nondimensionalize{\thermconductivity}{\freestream{\thermconductivity}}
    ,\;
    \nondimensionalize{\viscosity}{\freestream{\viscosity}}
    ,\;
    \nondimensionalize{\totale}{\freestream{\soundspeed}^2}
    ,\;
    \nondimensionalize{\totalh}{\freestream{\soundspeed}^2}
    ,\;
    \massprod = \frac{\dimval{\dot{\omega}}_\speciess}{\freestream{\density}\freestream{\soundspeed}/\valref{L}}
    \label{eq:nondimensionalization}
\end{alignedat}
\end{equation}

The definitions in \cref{eq:definitionseh} are used for the nondimensional total energy, \totale, internal enthalpy, \internalh, and total enthalpy, \totalh, respectively. 

\begin{equation}
    \totale = \frac{\internale}{\Eckert} + \frac{1}{2}\norm{\velocity}^2
    \;,\quad
    \internalh = \internale + \frac{\pressure}{\density}
    \;,\quad
    \totalh = \totale + \frac{\pressure}{\density},
    \label{eq:definitionseh}
\end{equation}
where the nondimensional internal energy is denoted \internale. The stress tensor, \viscstresstensor, and heat flux, \heatflux, are computed as in \cref{eq:definitionstress,eq:definitionheatflux}, respectively.

\begin{align}
    \viscstresstensor
    &= \frac{\viscosity}{\Reynolds_\freest} \left( \grad{\velocity} + \transposep{\grad{\velocity}} - \left(\divg{\velocity}\right)\eyematrix \right)
    \label{eq:definitionstress} \\
    \heatflux
    &= - \frac{\thermconductivity}{\Reynolds_\freest \Prandtl_\freest \Eckert_\freest}
    \grad{\temperature}
       + \sumspecies \densityspecies \speciesh \diffvelocity
    \label{eq:definitionheatflux}
\end{align}

The second term on the right-hand side of equation \cref{eq:definitionheatflux} is included only in the case of finite-rate chemistry (see \cref{sub:finiteratechem}). 
The nondimensional Reynolds number, $\Reynolds_\freest$, and Prandtl number, $\Prandtl_\freest$, are defined in the free stream at the domain inlet. The Eckert number, $\Eckert_\freest$, is also computed at the free stream and is equal to one by design.
These nondimensional quantities are defined as
\begin{equation}
    \Reynolds_\freest = \frac{\freestream{\density}\freestream{\soundspeed}\valref{L}}{\freestream{\viscosity}}
    \;,\quad
    \Prandtl_\freest  = \frac{\freestream{\viscosity}\freestream{\heatp}}{\freestream{\thermconductivity}}
    \;,\quad
    \Eckert_\freest   = \frac{\freestream{\soundspeed}^2}{\freestream{\heatp}\valref{\temperature}}
    .
    \label{eq:nondimnumbers}
\end{equation}
The thermodynamic properties and equation of state, as well as the transport properties and the chemical kinetics terms depend on the modeling choices and are discussed in detail in the following \cref{sub:thermochem}.
\subsection{Thermodynamic and chemical models}
\label{sub:thermochem}

A variety of models with different levels of complexity are implemented in the solver used in this work.
In the following, we discuss the details of each modeling approach and the implications of the underlying assumptions on the closure of the governing equations.

\subsubsection{Multi-component gas mixtures with variable composition}
\label{sub:gasmixtures}

In the general case, a reacting gas in a high-enthalpy flow is considered a multi-component mixture that consists of a set of species \speciesset, interacting via a defined network of reactions, as seen in \cref{sub:finiteratechem}. The variation of the species fractions makes the chemical reaction and diffusion terms in the governing equations significant, hence they need to be modeled using various assumptions.

For a multi-component gas mixture, the global mixture properties are derived from the species properties based on, 
\begin{equation}
    \density = \sumspecies \densityspecies
    \;,\quad
    \internale = \sumspecies \speciesmassfrac \speciese
    \;,\quad
    \internalh = \sumspecies \speciesmassfrac \speciesh \;,
    \label{eq:mixsums}
\end{equation}
where the species mass fraction is $\speciesmassfrac = \frac{\densityspecies}{\density}$, with $\sumspecies \speciesmassfrac = 1$. The mixture's thermodynamic and transport properties depend, in general, on any two thermodynamic properties, for example the temperature and the pressure, which define the thermodynamic state of the mixture, and on its composition. The composition is generally an independent variable. It can potentially become dependent on other thermodynamic quantities for the special cases seen in \cref{sub:frozenLTE}.

The individual species properties are accurately computed using kinetic theory and statistical mechanics \citep{scoggins_2014_mpp, Scoggins2017}.
The thermochemical library \acrshort{Mutation++} \cite{Scoggins2020} is used to compute thermodynamic and transport properties at different conditions.

\paragraph{Finite-rate chemistry and \texorpdfstring{\CNEQ}{CNEQ}}
\label{sub:finiteratechem}

When finite-rate chemistry cannot be neglected, the species mass production rate, \massprod, and the species diffusion velocities, \diffvelocity, need to be modeled. These terms are described in this section.

In a general case, a set of reactions, \reactionsset, is considered depending on the mixture in question. 
Each reaction, \reactionr\, is characterized by a reaction rate, \reactionrate, which is computed by the forward rate, \forwardrate, and the backward rate, \backwardrate. These, in turn, are obtained according to experimentally or theoretically calibrated Arrhenius formulas in the form $\forwardrate = C_\reactionr \temperature^{n_\reactionr} \exp{\left({\temperature_a}_\reactionr/\temperature\right)}$, and $\backwardrate = \forwardrate/{\equilconst{(\temperature)}}$, where \equilconst\ is the reaction equilibrium constant at specific conditions.
This description is given in \cref{eq:reaction} for a generic reaction. 

\begin{equation}
    \text{Reaction (r):}
    \quad\quad
    \sumspecies \nu'_{\reactionr,\speciess} S_\speciess
    \xrightleftharpoons[\backwardrate]{\forwardrate} 
    \sumspecies \nu''_{\reactionr,\speciess} S_\speciess
    \quad\quad
    \label{eq:reaction}
\end{equation}
The net reaction rate is then given by
\begin{equation}
    \reactionrate = \left[ \forwardrate \Pi_\speciess \left( \frac{\densityspecies}{\mwspecies} \right)^{\nu'_{\reactionr,\speciess}}
    - \backwardrate \Pi_i \left( \frac{\densityspecies}{\mwspecies} \right)^{\nu''_{\reactionr,\speciess}} \right] 
    \cdot \sumspecies \left( Z_{\reactionr,\speciess} \frac{\densityspecies}{\mwspecies} \right),
    \label{eq:reactionrate}
\end{equation}
where the species molar mass is \mwspecies\ and the efficiency of species \speciess\ as a third-body in reaction \reactionr\ is $ Z_{\reactionr,\speciess}$.
The net mass production rates for species \speciess\ from all reactions are obtained as 
\begin{equation}
    \massprod = \mwspecies \sumreactions \left( \nu''_{\reactionr,\speciess} - \nu'_{\reactionr,\speciess} \right) \reactionrate.
    \label{eq:reactionmass}
\end{equation}
The reader is referred to the description provided in \cite{Scoggins2017,Scoggins2020}, on which the \acrshort{Mutation++} library is based, for further details.

The diffusion flux, $\vec{J}_\speciess = \densityspecies \diffvelocity$ appearing in \cref{eq:speciesmass} also needs to be modeled in this case. Following the description in \cite{Marxen2013}, and under the same assumptions, neglecting thermal and barodiffusion, the diffusion driving force for each species, $\vec{d}_\speciess$, reduces to its molar fraction gradient,
\begin{equation}
    \vec{d}_\speciess = \grad{\speciesmolefrac}.
    \label{eq:drivingforces}
\end{equation}
In the general case, the diffusion velocity for each species \speciess, $\diffvelocity$, is then the solution of the Stefan-Maxwell linear system of equations for a multi-component mixture,
\begin{equation}
    \left\{\vphantom{\frac{1}{1}}\right.
    \sum_{\subs{i}\in\speciesset} \tens{Q}_{\speciess,\subs{i}} \vec{V}_\subs{i} = - \vec{d}_\subs{s}
    \left.\vphantom{\frac{1}{1}}\right\}
    , \quad \forall \; \subs{s} \in \speciesset
    \label{eq:stefanmaxwell}
\end{equation}
where, $\tens{Q}_{\speciess,\subs{i}}$ is the Stefan-Maxwell interaction coefficient for the pair of species $\speciess$ and $\subs{i}$.

While solving the system in \cref{eq:stefanmaxwell} is the most accurate way to calculate the diffusion fluxes in a multi-component gas mixture, simpler formulations are found in the literature under the same assumptions \cite{ramshaw1990,ern1994}. A simpler expression based on Fick's diffusion model (applicable to binary mixtures) with a mass correction is given below in \cref{eq:ramshaw},

\begin{equation}
    \vec{J}_\speciess = -c \mwspecies D_\speciess \grad{\speciesmassfrac} + c \speciesmassfrac \sum_{\subs{i} \in \mathrm{S}} \mw_\subs{i} D_\subs{i} \nabla Y_\subs{i}.
    \label{eq:ramshaw}
\end{equation}

Here, $c = \sumspecies (\densityspecies / \mwspecies)$ and $D_\speciess$ is the averaged diffusion coefficient for species \speciess, defined as
\begin{equation}
    D_\speciess = \frac{1 - \speciesmolefrac}{\sum_{\subs{r} \neq \speciess} X_\subs{r}/D_{\speciess, \subs{r}}}.
    \label{eq:speciesavgdiffusioncoef}
\end{equation}
In lieu of solving a linear system of equations of size $N_\speciess$, using this model, the diffusion flux is computed from an algebraic equation, at a significantly reduced computational cost. Both implementations are available in the solver presented here, and the accuracy of the simplified approach is verified for practical cases. The constraints seen in \cref{eq:sumomegaanddiff} have to be respected for the kinetic and diffusive terms, hence
\begin{equation}
    \sumspecies \massprod = 0
    \;,\quad
    \sumspecies \densityspecies \diffvelocity = 0.
    \label{eq:sumomegaanddiff}
\end{equation}
The heat flux takes the original form shown in \cref{eq:definitionheatflux}, where the thermal conductivity is replaced by its frozen composition value. The diffusive heat flux is then taken into account explicitly through the second term in that definition.
The remaining relevant thermodynamic and transport properties are given by general relations that are implemented in the \acrshort{Mutation++} library.

\paragraph{Special cases: frozen and equilibrium flow}
\label{sub:frozenLTE}
 
There exist two extreme cases for a multi-component mixture, where its composition becomes irrelevant.%
\begin{itemize}
    \item \textbf{Frozen composition}, when the reaction time scale is too long compared to the flow time scale, hence the composition of the flow is generally constant in time.
    \item \textbf{\Acrfull{LTE}}, when the reaction time scale is short enough for the reactions to bring the mixture to a state of local equilibrium, where the composition is identical to its equilibrium composition that can be computed from thermodynamic relations by maximizing the system entropy or, equivalently, minimizing the Gibbs free energy.
\end{itemize}

In both cases of frozen composition and \LTE, the mixture composition becomes a dependent variable, directly computed for a given thermodynamic state. Therefore, the species mass conservation equations, \cref{eq:speciesmass}, and the corresponding reaction and diffusion terms are not needed. The heat flux takes the simplified form shown in \cref{eq:heatfluxeq}, 
\begin{equation}
    \heatflux
    = - \frac{\thermconductivity}{\Reynolds_\freest \Prandtl_\freest \Eckert_\freest}
    \grad{\temperature},
    \label{eq:heatfluxeq}
\end{equation}
where the thermal conductivity is computed at either frozen or \LTE\ conditions. In the latter case, the diffusive effect at equilibrium is included in its value.

In situations between these extremes, finite-rate chemistry needs to be tracked as described in \cref{sub:finiteratechem} and the mixture composition changes in space and time.

\subsubsection{Calorically and thermally perfect gases}
\label{sub:perfectgas}

When the gas is assumed to be calorically or thermally perfect, the composition becomes irrelevant. The chemical diffusion and reaction terms in the governing equations vanish, and the thermodynamics and transport properties become simple functions of the thermodynamic state of the gas, resulting in a nondimensional equation of state given by
\begin{equation}
    \pressure
    = \density \frac{\gammaratio - 1}{\gammaratio \Eckert_\freest} \temperature. 
    \label{eq:stateequation}
\end{equation}
It should be noted that this formulation is valid in this form only for the \CPG\ and \TPG\ models. In the case of reacting mixtures with variable composition, the right hand side needs to be scaled by a factor accounting for the change of gas constant and depends on the composition.

In the case of a \CPG, the thermodynamic properties, such as the specific heat, are considered constant. The internal energy and enthalpy thus become linear functions of temperature, in nondimensional form $\gammaratio_\freest\internale = \temperature$, which is derived from the dimensional form $\dimval{\internalh} = \freestream{\heatp} \dimval{\temperature}$. Nondimensionally, $\thermconductivity=\viscosity$.

In the case of a \TPG, the thermodynamic properties are generally functions of temperature. Since the specific heat is now itself a function of temperature, the internal energy and enthalpy become nonlinear functions of temperature. The relations used are based on the work presented in \cite{Malik1991, Marxen2011}. The specific heat is given as 
\begin{equation}
    \dimval{\heatp}\left(\dimval{\temperature}\right)
    = \dimval{\heatp}^0 \left(1 + \frac{\freestream{\gammaratio} - 1}{\freestream{\gammaratio}}
    \left(\frac{\dimval{\Theta}}{\dimval{\temperature}}\right)^2
    \frac{e^{\dimval{\Theta}/\dimval{\temperature}}}{e^{\dimval{\Theta}/\dimval{\temperature}} - 1}\right),
    \label{eq:cpfunction}
\end{equation}
where $\dimval{\Theta} = \SI{3055}{K}$ and $\dimval{\heatp}^0$ such that $\dimval{\heatp}\left(\freestream{\temperature}\right) = \freestream{\heatp}$.
The transport properties (viscosity and thermal conductivity) can be constant or follow analytical expressions as functions of temperature.

In both the \CPG\ and \TPG\ assumptions, the viscosity is computed using Sutherland's law,
\begin{equation}
    \dimval{\viscosity}\left(\dimval{\temperature}\right)
    = \dimval{C}_1 \frac{\dimval{\temperature}^{3/2}}{\dimval{\temperature} + \dimval{T}_S},
    \label{eq:sutherland}
\end{equation}
where $\dimval{C}_1 = \SI{1.458e-6}{kgm^{-1}s^{-1}K^{-1/2}}$ and $\dimval{\temperature}_S = \SI{110.4}{K}$. 

In addition, in both cases the heat flux takes the simplified form seen in \cref{eq:heatfluxeq} as the diffusive heat flux vanishes. In \CPG, the equilibrium conductivity is assumed constant, $\dimval{\thermconductivity} = \freestream{\thermconductivity}$, while $\thermconductivity$ can be computed as in \cref{eq:nondimensionalization}. Whereas, in \TPG, the equilibrium thermal conductivity is approximated by Keyes' law, 
\begin{equation}
    \dimval{\thermconductivity}\left(\dimval{\temperature}\right)
    = \dimval{C}_2
    \frac{\dimval{\temperature}^{1/2}}
    {1 + \left( \dimval{C}_3 / \dimval{\temperature}\right) 10^{-\dimval{C}_4 / \dimval{\temperature}}},
    \label{eq:keyes}
\end{equation}
with $\dimval{C}_2 = \SI{2.646e-3}{Wm^{-1}K^{-3/2}}$, $\dimval{C}_3 = \SI{245.4}{K}$, $\dimval{C}_4 = \SI{12}{K}$.

\section{Numerical framework}\label{sec:numerics}

The following section outlines the numerical framework of the Navier-Stokes solver and the implementation of the models described so far.

\subsection{Discretization in space and time}
\label{sub:numdiscretization}

The computational tools developed are based on the original Navier-Stokes solver developed by \citeauthor{nagarajan2004_thesis} \citep{nagarajan2003, nagarajan2004_thesis}. The code has been applied to \DNS\ and \LES\ studies of flat-plate boundary layers in the subsonic \cite{sayadi2013} and the supersonic \cite{Marxen2011,Marxen2013,Marxen2014,Marxen2014a} regime.

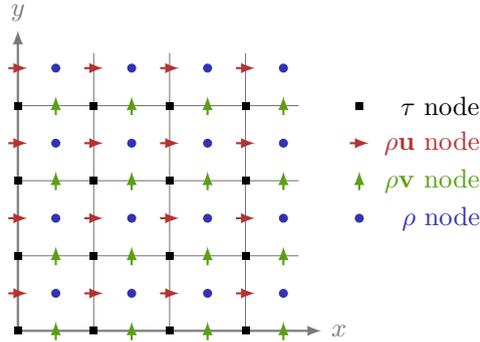
\begin{figure}
    \centering
    \input{Figures/Fig1.tex}
    \caption{Schematic of a staggered grid in two dimensions. The cell-centered $\density$ nodes are where scalar variables are stored and computed, while the interface nodes $\density u$ and $\density v$ are where the streamwise and wall-normal velocities, fluxes, and gradients are stored and computed. The $\tau$ nodes are the grid nodes where the grid is generated, and where the solution is interpolated for post-processing and presentation. Extension to three dimensions is trivial, with an additional $\density w$ node at the interface in the page-normal direction, for the spanwise velocities, fluxes, and gradients.}
    \label{fig:staggeredgrid}
\end{figure}

Space discretization is done using fourth-order or sixth-order compact finite differences \cite{Lele_1992_compactFD}. The governing equations are formulated in curvilinear coordinates and solved on a staggered grid, as seen in \cref{fig:staggeredgrid}. A detailed discussion of the curvilinear transformation is found in \cite{nagarajan2004_thesis}.

Time discretization is done using explicit third-order or fourth-order Runge-Kutta schemes. Specifically, the low-memory two-register RK3 scheme, the three-register \TVD\ RK3 and RK4 \citep{carpenter_1994_rk_schemes,kennedy_2000_rk_schemes} are implemented.

\subsection{Boundary conditions}
\label{sub:numbcs}

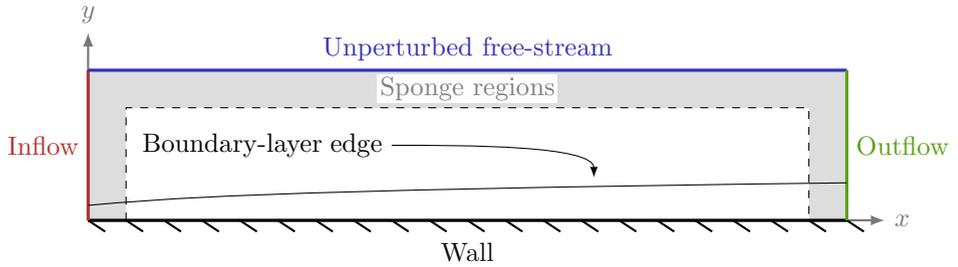
\begin{figure}
    \centering
    \input{Figures/Fig2.tex}
    \caption{Schematic of a generic computational domain. Inflow and outflow can be enforced with reference solutions. The same approach is used at the free-stream. The wall is non-catalytic, either isothermal or adiabatic. A variety of forced boundary conditions can be applied locally on the wall.}
    \label{fig:domain}
\end{figure}

A general sketch of the domain of interest is presented in \cref{fig:domain}. A reference solution (typically, a self-similar boundary layer) is prescribed in the sponge regions, and the time-advanced solution is forced towards that reference by incorporating damping source terms in the right-hand side of the equations, as seen in \cref{eq:sponge}.
\begin{equation}
    \pd{\statevec}{\timevar} = \fluxvec\left(\statevec\right) - \sponge\left(\spacevec\right) \left(\statevec - \statevec_\subsref \right)
    \label{eq:sponge}
\end{equation}
The sponge parameter $\sponge\left(\spacevec\right)$ is a smooth third-order polynomial function of space, vanishing inside the domain and reaching a high value near the boundaries which is selected empirically. The sponge can be completely omitted at the inflow in cases where it is not necessary, such as hypersonic flow where upstream-traveling disturbances are minimal. Flow perturbations are also forced to vanish inside the sponge regions. The wall boundary is either adiabatic or isothermal, while no slip and no catalysis is permitted for the reacting cases. Other boundary conditions, such as blowing and suction at the wall, isolated roughness geometries, or jet injections, are also implemented and available.

\subsection{Shock-capturing scheme}
\label{sub:numshock}
In the presence of shocks and discontinuities special treatment is necessary. Out of the variety of approaches that exist, we opted for a numerical scheme based on artificial molecular and bulk viscosity. This scheme has previously been implemented and verified for shock capturing in hypersonic flow simulations. Further details can be found in \cite{kawai2008_diffusivity,kawai2010_diffusivity,Mani_2009_diffusivity,fiorina2007_diffusivityspecies}.

\subsection{Coupling with the \texorpdfstring{\acrshort{Mutation++}}{Mutation++} library}
\label{sub:mppcoupling}

The thermodynamic and transport properties, and the source terms for chemical kinetics models, are extracted from the \acrshort{Mutation++} library \cite{Scoggins2020}. The library, written in \texttt{C++}, is coupled with the solver, written in \texttt{Fortran 95}, using a wrapper interface that facilitates the library function calls by implementing them as functions and subroutines in the solver.
An on-the-fly communication between the solver and the library is necessary, with constant evaluation of the thermochemical and transport properties at each grid point, given the local state vector. Quantities are therefore exchanged between the solver and the library for each grid point, at each time step. These function evaluation calls add a significant computational overhead to the solver, since each state evaluation involves the iterative solution of nonlinear equations. Concepts and techniques that would accelerate these evaluations are investigated as part of ongoing work. The library offers valuable modularity, decoupling the thermochemical model from the core of the solver, and offering multiple capabilities to change gas mixtures, databases, and models.

\subsubsection{Comparison between perfect-gas and real-gas models}

Benchmark simulations have been run to compare the accuracy and computational cost of the various thermochemical models implemented in the solver. The flexibility and extended validity of the use of \acrshort{Mutation++} comes with a large computational cost. The total simulation time is increased by more than one order of magnitude when using the \CNEQ\ model compared to the \CPG\ model, and by a factor of about five compared to the \TPG\ model. A reduction of about 40\% in computational cost can be achieved using an algebraic diffusion model, as discussed in \cref{sub:finiteratechem}, compared to the Stefan-Maxwell diffusion model, with minimal impact on the accuracy for the cases investigated in this work.

\section{Results}\label{sec:results}
In this section, a set of cases are investigated in order to verify the methodology presented so far and to investigate the effect of gas model selection on the resulting flow in different configurations. These cases include adiabatic and isothermal $Ma=10$ laminar boundary layers, forced with Tollmien-Schlichting waves (see \cref{table:condition_olaf}), laminar \SBLI\ at $\Mach=2$ and $\Mach=5.92$, and a jet in hypersonic crossflow at $\Mach=5$.

\subsection{Hypersonic flat-plate boundary layers}\label{sub:hypersonic}
 
A set of two hypersonic boundary layers in Earth's atmosphere at $Ma = 10$, based on \citeauthor{Marxen2013} \citep{Marxen2013,Marxen2014}, are considered here to verify the implementation of the numerical method presented in \cref{sec:numerics}. First, the steady state solution and then the growth rate of forced Tollmien-Schlichting instabilities inside the two-dimensional boundary layer are compared.
The main configurations considered are: (i) case I (isothermal), and (ii) case A (adiabatic), referring to the boundary condition imposed on the wall. The setup parameters and freestream conditions for both cases are presented in \cref{table:condition_olaf}, where $\omega$ and $A$ are related to the forced perturbation and explained in \cref{sub:forcing}.

\begin{table}[tbph]
\begin{center}
\begin{tabular}{ |c||c|c| } 
    \hline
    & Isothermal Case (I) & Adiabatic Case (A) \\
    \hline
    $ \Mach_\infty $ & \multicolumn{2}{c|}{10.0}\\ 
    \hline
    $ \Reynolds_\infty $ & \multicolumn{2}{c|}{$10^5$}\\ 
    \hline
    $ \freestream{\temperature} [K] $ & 278  & 350  \\ 
    \hline
    $ \freestream{\pressure} [Pa] $ & 4135 & 3596 \\ 
    \hline
    $ \temperature_\subs{wall}/\freestream{\temperature} $ & 4.31 & - \\ 
    \hline
    $ \omega $ & 45 & 34 \\ 
    \hline
    $ A/\Mach_\infty $ &  \multicolumn{2}{c|}{$10^{-3}$} \\ 
    \hline
\end{tabular}
\caption{Thermodynamic and freestream conditions for the $\Mach=10$ hypersonic boundary layers investigated in this study (adapted from \cite{Marxen2013}).}
\label{table:condition_olaf}
\end{center}
\end{table} 

The computational domain considered here has a finer resolution than in the previous studies. The nondimensional grid size in the streamwise direction is $\Delta x = 0.075$, simulating a domain from $x_0 = 14.0$ to $x_1 = 86.0$ using 960 grid points. The sponge regions extend for $5$ and $15$ nondimensional units at the inflow and outflow, respectively. In the wall-normal direction, 211 grid points are used, clustered near the wall using \cref{eq:stretching}, with the last 26 points in the free stream used in the sponge region. \cref{table:mesh_bl} summarizes the details of the computational domain as well as the resulting resolution. 

\begin{equation}
    \label{eq:stretching}
    y(m) = y_0 + (y_1 - y_0)\left((1-\kappa_y)\left( \frac{m-1}{NY-1}\right)^3 + \kappa_y \left(\frac{m-1}{NY-1}\right)\right), \; m \in \left[ 1,NY \right]
\end{equation}

\begin{table}[tbph]
\begin{center}
\begin{tabular}{ |r|| c c | c c | c c | c c | } 
 \hline
 Case & $x_0$ & $x_1$ & $y_0$ & $y_1$ & $\kappa_x$ & $\kappa_y$ & $N_x$ & $N_y$ \\ 
 \hline
 \hline
 Case A & $14.0$ & $86.0$ & $0$ & $1.6$ & $1.0$ & $0.15 $ & 960 & 211 \\
  \hline
 Case I & $14.0$ & $86.0$ & $0$ & $1.6$ & $1.0$ & $0.15 $ & 960 & 211 \\
  \hline
\end{tabular}
\caption{Mesh configuration for the hypersonic boundary layer simulations}
\label{table:mesh_bl}
\end{center}
\end{table}

All three thermochemical models are systematically tested on both cases. The \CNEQ\ model uses a 5-component air mixture composed of $\text{N}_2, \text{O}_2, \text{N}, \text{O}$, and $\text{NO}$.

\subsubsection{Disturbance forcing}\label{sub:forcing}

Tollmien-Schlichting waves are harmonically forced on the converged steady-state solution of the unforced equations using blowing and suction on a strip extending from $x=19.3$ to $x=20.7$. For case A, the wall condition is switched to isothermal for the perturbations, assuming only small deviations from the steady-state temperature. The disturbance has a nondimensional forcing frequency $\omega$ given in \cref{eq:disturbance}.

\begin{equation}
    \omega = F \Mach_{\infty}^2 \Reynolds_{\infty}, \quad \text{with} \quad F = 2\pi\tilde{f} \frac{\freestream{\viscosity}}{\freestream{\density} \freestream{u}^2}
    \label{eq:disturbance}
\end{equation}

Here $\tilde{f}$ is the dimensional frequency of the disturbance. The forcing amplitude is small enough to ensure that the disturbance evolution is linear, i.e. $A = 10^{-3} \Mach_\infty$.

The simulation is advanced until transient effects are advected out of the domain and a time-periodic state is achieved. Flow snapshots are collected over two forcing periods and the Fourier transformation is applied, giving a complex Fourier coefficient $\Hat{\phi}_\text{j}$ for a given harmonic $j$ and a given quantity $\phi \in [\density, u, v, w, \temperature, \ldots]$. Only the results for the first harmonic are presented in \cref{caseI} and \cref{caseA}. The streamwise disturbance amplification is computed using the wall-normal maxima of the amplitudes of the disturbance quantity $\hat{\phi}_1$,
\begin{equation}
    \hat{\phi}_1^{max}(x) = \max_{y} (\lvert \hat{\phi}_1(x,y)\rvert)
\end{equation}
and the normalized growth rate $\alpha_{i,1}$ is computed as follows, for example for the wall-normal velocity, i.e  for $\phi=v$, it is
\begin{equation}
    \alpha_{i,1} (x) = \frac{1}{\hat{v}_1^{max}} \frac{\partial \hat{v}_1^{max}}{\partial x} \frac{\Reynolds_x}{\Reynolds_\infty} \text{.}
\end{equation}

In all the above, sponge regions are not considered in the calculations. In the following analysis, the streamwise varying Reynolds number is defined as $\Reynolds_x=\sqrt{x \Reynolds_\infty \Mach_\infty}$.

\subsubsection{Isothermal wall -- Case I}\label{caseI}

\begin{figure}[tbph]
    \centering
    \begin{subfigure}[t]{0.49\textwidth} 
        \centering 
        \includegraphics[width=\textwidth]{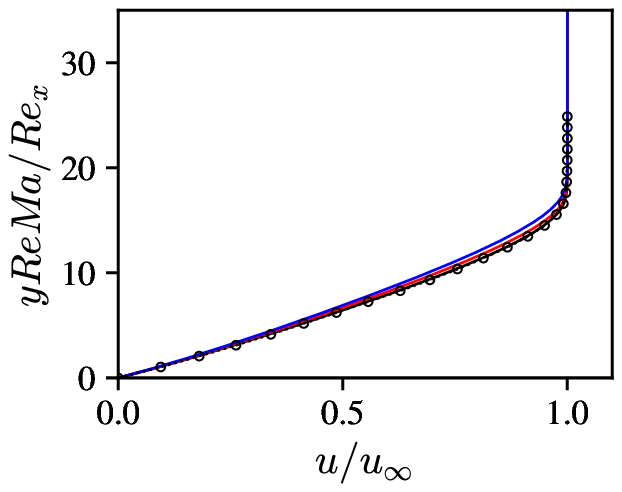}
        \caption{}\label{fig:vel_I}
    \end{subfigure}
    \hfill
    \begin{subfigure}[t]{0.49\textwidth}
        \centering 
        \includegraphics[width=\textwidth]{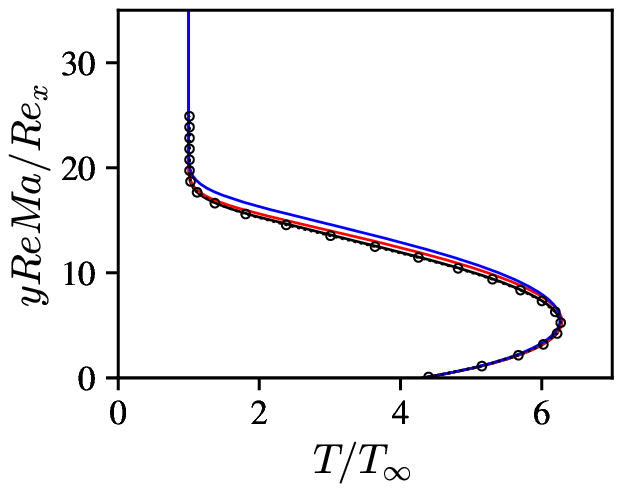}
        \caption{}\label{fig:temp_I}
    \end{subfigure}
    \caption{(a) Streamwise velocity, $u$, and (b) temperature, $\temperature$, profiles in the boundary layer for case I at $\Reynolds_x = 2000$. Current results (solid lines) are compared to \citet{Marxen2013} (dotted lines with symbols) for different gas models: \TPG\ (black), \LTE\ (blue), \CNEQ\ (red).}\label{fig:profiles_caseI}
\end{figure}

The mean streamwise velocity and temperature profiles at a streamwise Reynolds number location $\Reynolds_x = 2000$ for case I are presented in \cref{fig:profiles_caseI}, compared with the results of \citet{Marxen2013}, where only the \TPG\ model was investigated. This figure shows both \TPG\ profiles to be in perfect agreement as expected. In addition, both the \LTE\ and \CNEQ\ models show only minor differences compared to the \TPG\ solution. This is expected as the isothermal condition induces a maximum temperature of $\temperature_\subs{max} \approx 1740 K$ in the boundary layer. Below 2000K air chemistry is known to be mostly irrelevant \cite{Anderson2019} with almost negligible dissociation. Hence, the thermally perfect gas assumption holds reasonably well, resulting in similar behavior regardless of the model selected.

\begin{figure}[tbph]
    \centering
    \begin{subfigure}[t]{0.49\textwidth} 
        \centering
        \includegraphics[width=\textwidth]{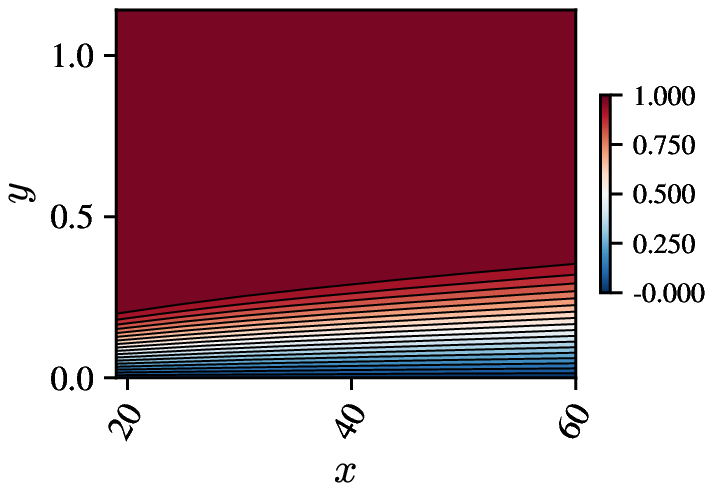}
        \caption{}\label{fig:contour_vel_I}
    \end{subfigure}
    \hfill
    \begin{subfigure}[t]{0.49\textwidth}
        \centering
        \includegraphics[width=\textwidth]{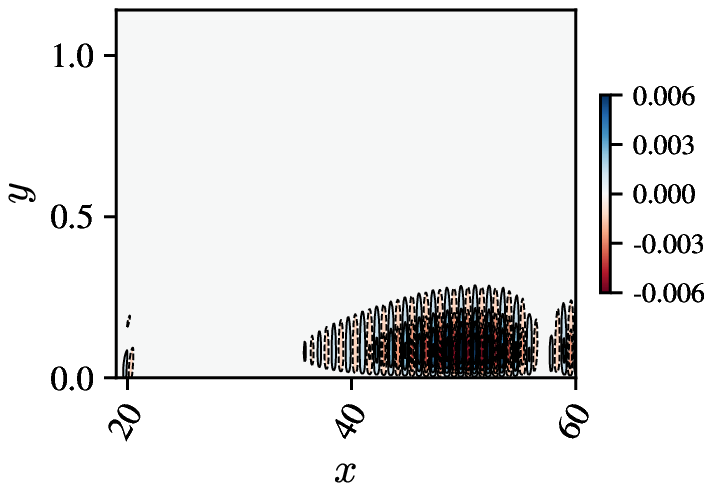}
        \caption{}\label{fig:contour_v'_I}
    \end{subfigure}
    \begin{subfigure}[t]{0.55\textwidth}
        \centering
        \includegraphics[width=\textwidth]{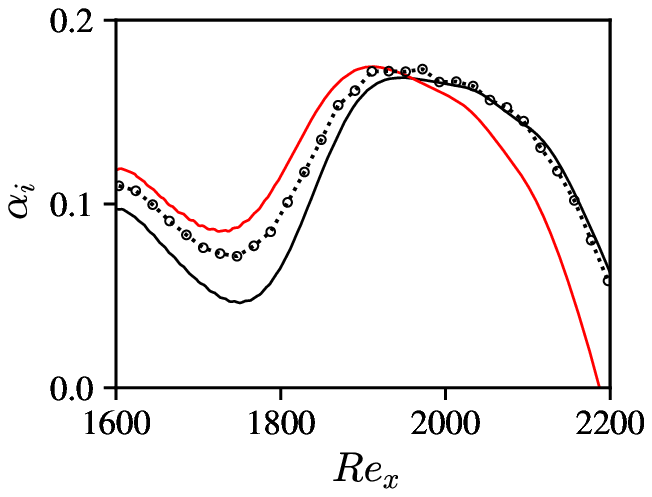}
        \caption{}
        \label{fig:growth_caseI}
    \end{subfigure}
    \caption{(a) Mean streamwise velocity field, $u$, and (b) wall-normal perturbation velocity field, $v'$, for case I. (c) Streamwise growth rate, $\alpha_i$, of the linear perturbation in the isothermal boundary layer with respect to $\Reynolds_x$. Current results (solid lines) are compared to \citet{Marxen2013} (dotted lines with symbols) for different gas models: \TPG\ (black), \CNEQ\ (red).}\label{fig:perturbed_caseI}
\end{figure}

In \cref{fig:perturbed_caseI} we see the baseflow solution for the streamwise velocity, in \cref{fig:contour_vel_I}, and a snapshot of the evolution of the forced perturbation, in \cref{fig:contour_v'_I}. Finally, \cref{fig:growth_caseI} shows the growth rate $\alpha_i$ of the Tollmien-Schlichting waves in this case. While some differences are observed upstream, perfect agreement is found downstream with the results of \citet{Marxen2013} for the \TPG\ model. The current simulations are better resolved than those of \citet{Marxen2013}, and the improved resolution has been found to explain the more accurate prediction and the disappearance of small oscillations upstream. Interestingly, despite the \CNEQ\ baseflow being almost identical to that of the \TPG\ model, the instability grows earlier in the boundary layer and to a higher amplitude compared to the case modeled using a \TPG\ assumption. Thus, finite-rate chemistry alters the growth and decay of the perturbations directly, even though it has a negligible effect in the baseflow solution, potentially altering the stability and transition behavior without an obvious effect on the baseflow.

\subsubsection{Adiabatic wall -- Case A}\label{caseA}

In the adiabatic case, a higher freestream temperature is imposed by design to promote \CNEQ\ effects.

\begin{figure}[tbph]
    \centering
    \begin{subfigure}[t]{0.49\textwidth} 
        \centering \includegraphics[width=\textwidth]{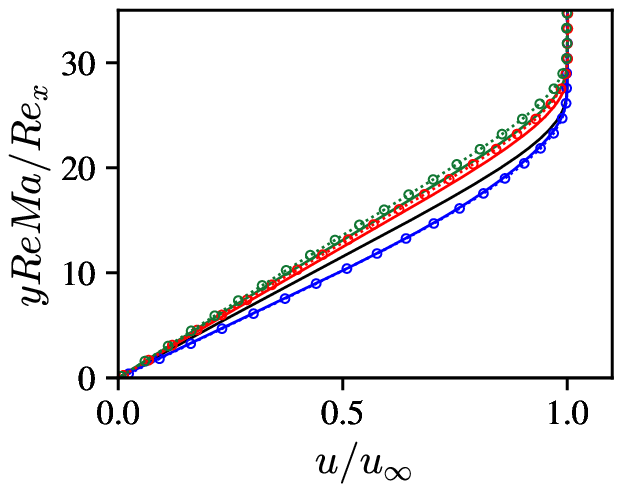}
        \caption{}\label{fig:vel_A}
    \end{subfigure}
    \hfill
    \begin{subfigure}[t]{0.49\textwidth}
        \centering \includegraphics[width=\textwidth]{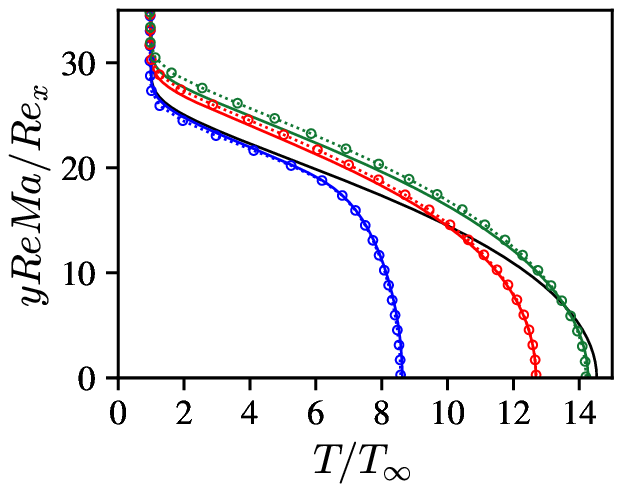}
        \caption{}\label{fig:temp_A}
    \end{subfigure}
    \begin{subfigure}[t]{0.49\textwidth}
        \centering \includegraphics[width=\textwidth]{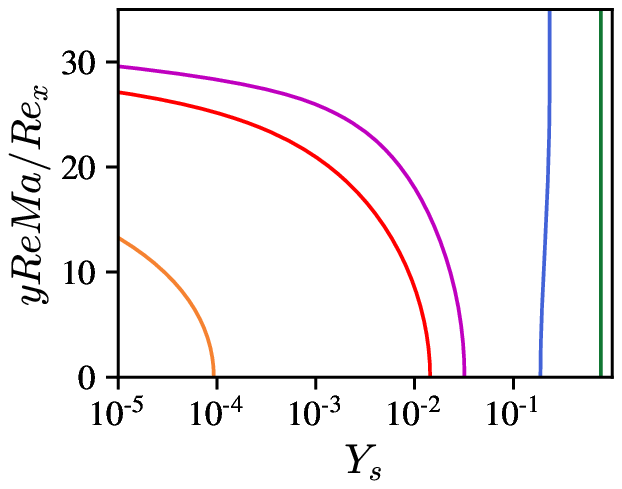}
        \caption{}\label{fig:species_profile}
    \end{subfigure}
    \caption{(a) Streamwise velocity, $u$, and (b) temperature, $\temperature$, profiles in the boundary layer for case A at $\Reynolds_x = 2000$. Current results (solid lines) are compared to \citet{Marxen2013} (dotted lines with symbols) for different gas models: \TPG\ (black), \LTE\ (blue), \CNEQ\ (red), frozen chemistry (green). (c) Species mass fractions at $\Reynolds_x = 2000$. From left to right : $N$, $NO$, $O$, $O_2$, $N_2$.}\label{fig:profiles_caseA}
\end{figure}

The mean flow and temperature profiles at a streamwise Reynolds number location $\Reynolds_x = 2000$ are presented in \cref{fig:profiles_caseA} and compared to the results of \citet{Marxen2013}. Similar to the isothermal case, practically perfect agreement is found for all models compared to the previous results. However, in this case the baseflows differ significantly depending on the model used for the gas. Wall temperature decreases significantly from a frozen-chemistry (fixed-composition) assumption to a \CNEQ\ model to an \LTE\ model. These results indicate that the \CNEQ\ effects in the flow are significant in this case and need to be accounted for to achieve accurate baseflow predictions.

Due to the near-wall temperature approaching $\temperature_\subs{wall}\approx 4900K$ near the inflow, $\textrm{N}_2$ and $\textrm{O}_2$ rapidly start to dissociate to $\textrm{N},\textrm{O},\textrm{NO}$ through endothermic chemical reactions. This is illustrated in \cref{fig:species_profile}, presenting all the mass fraction profiles at the streamwise location where $\Reynolds_x = 2000$. Close to the wall, $\textrm{O}_2$ mass fraction decreases while $\textrm{O}$ and $\textrm{NO}$ are produced. To a smaller extend, $\textrm{N}$ is also created through $\textrm{N}_2$ dissociation. Moreover, the dissociated species concentrations ($\textrm{O},\textrm{NO},\textrm{N}$) build up as the species are also convected downstream while continuously being produced.
Consequently, the wall temperature decreases progressively along the streamwise direction due to cooling by endothermic dissociation.

\begin{figure}[tbph]
    \centering
    \begin{subfigure}[t]{0.49\textwidth} 
        \centering
        \includegraphics[width=\textwidth]{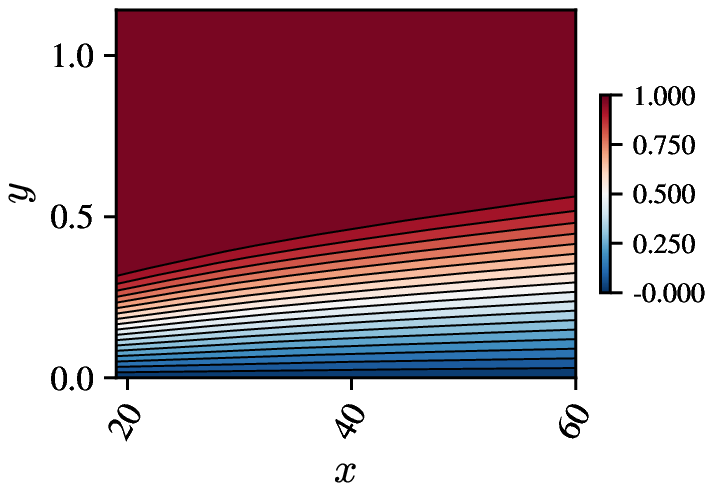}
        \caption{}
        \label{fig:contour_vel_A}
    \end{subfigure}
    \hfill
    \begin{subfigure}[t]{0.49\textwidth}
        \centering
        \includegraphics[width=\textwidth]{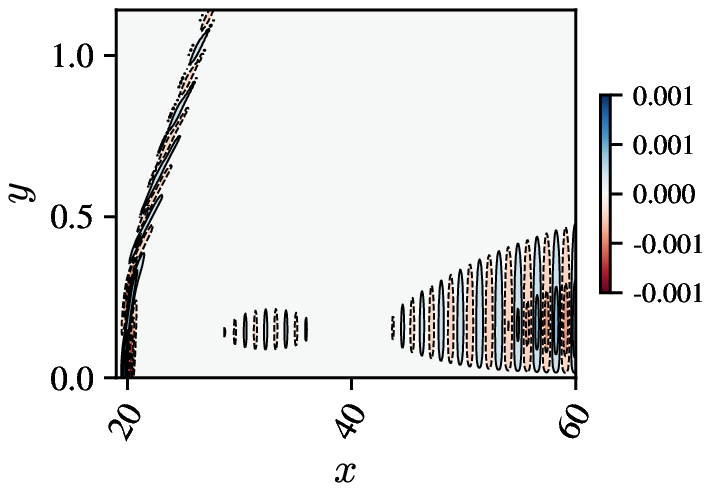}
        \caption{}
        \label{fig:contour_v'_A}
    \end{subfigure}
    \begin{subfigure}[t]{0.55\textwidth}
        \centering
        \includegraphics[width=\textwidth]{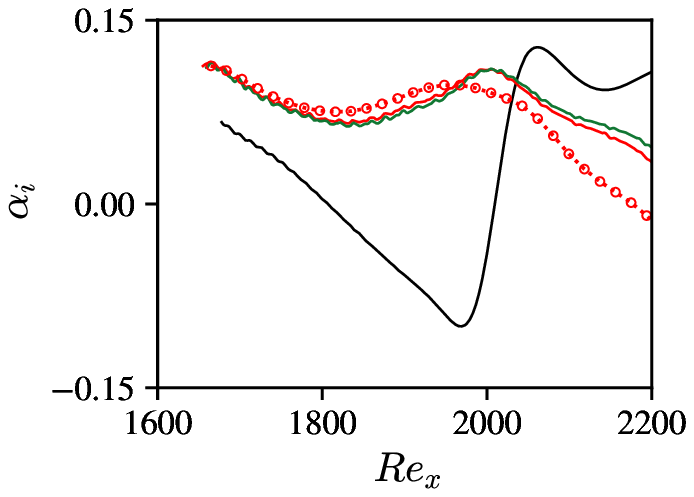}
        \caption{}
        \label{fig:growth_caseA}
    \end{subfigure}
    \caption{(a) Mean streamwise velocity field, $u$, and (b) wall-normal perturbation velocity field, $v'$, for case A. (c) Streamwise growth rate, $\alpha_i$, of the linear perturbation in the adiabatic boundary layer with respect to $\Reynolds_x$. Current results (solid lines) are compared to \citet{Marxen2013} (dotted lines with symbols) for different gas models: \TPG\ (black), \CNEQ\ with the Stefan-Maxwell multicomponent diffusion model (red), \CNEQ\ with Ramshaw's simplified diffusion model (green).}\label{fig:perturbed_caseA}
\end{figure}

In \cref{fig:perturbed_caseA} we see the baseflow solution for the streamwise velocity, in \cref{fig:contour_vel_A}, and a snapshot of the evolution of the forced perturbation, in \cref{fig:contour_v'_A}. \Cref{fig:growth_caseA} shows the corresponding growth-rates in the adiabatic case. Correct agreement is found for the \CNEQ\ model up to $\Reynolds_x = 1950$. Downstream, the computed growth rate differs slightly from the previous results but the overall trend is similar. This difference is explained by the evolution of the thermodynamic and kinetics models used in the corresponding libraries. The growth rate in the \TPG\ model is however noticeably different compared to the cases using the \CNEQ\ model, highlighting the inadequacy of such models in high-enthalpy cases.

In order to confirm the reduction in CPU time using Ramshaw's simplified algebraic diffusion model (see \cref{sub:finiteratechem}), with negligible impact on the accuracy of the results, the base flow and growth rates of Tollmien-Schlichting waves for case A are computed using equation \cref{eq:ramshaw}.
Base flows are practically identical in terms of all relevant variables. Similarly, as shown in \cref{fig:growth_caseA}, the growth rate is in almost perfect agreement with that computed using the Stefan-Maxwell model. A slight discrepancy is observed, which can be attributed to small changes in the diffusion fluxes. Therefore, \cref{eq:ramshaw} is a good compromise between performance and physical accuracy. The same model was also used in various numerical studies of hypersonic boundary layer with finite-rate chemistry \cite{passiatore2021_prf,direnzo2020_htr,direnzo2021_dns,direnzo2021_htr12}.

\subsection{\texorpdfstring{\Acrshort{2D} \acrlong{SBLI}}{2D shockwave--boundary-layer interaction}}\label{SBLI}
Supersonic and hypersonic flows over complex geometries usually present \SBLI. The large pressure gradient induced by the impinging shock may cause separation of the boundary layer with the occurrence of a recirculation bubble. This bubble can in turn change the stability characteristics of the flow on the vehicle surface. In the following section, we first validate the capability of the code to accurately simulate \SBLI\ against a benchmark shockwave laminar boundary layer experiment. Then, a higher Mach number case is designed, based on \cite{hildebrand2018}, with a high freestream temperature, to directly assess the effect of finite-rate chemistry compared to a perfect gas assumption.

\begin{table}[tbph]
\begin{center}
\begin{tabular}{ |l|| c c | c c |c c| } 
 \hline
 \SBLI\ Case & $x_0 - x_1$ & $y_0 - y_1$ & $\kappa_x$ & $\kappa_y$ & $N_x$ & $N_y$ \\ 
 \hline
$\Mach_\infty=2.0$  & $0.30 - 1.55 $ & $0.0 - 0.35$ & $1.0$ & $0.15 $ & 960 & 501 \\
$\Mach_\infty=5.92$ & $19.0 - 254.0$ & $0.0 - 36.0$ & $1.0$ & $0.15 $ & 960 & 450 \\
\hline
\end{tabular}
\caption{Mesh configuration for the two \SBLI\ cases investigated.}
\label{table:mesh_swbli}
\end{center}
\end{table}

\subsubsection{2D \texorpdfstring{\SBLI\ at $\Mach=2$}{SBLI at Ma = 2}}

The capability of the solver to correctly simulate compressible flows including shocks is verified using a benchmark \SBLI\ case, first investigated experimentally by \citet{hakkinen1959} and later numerically in several studies \cite{katzer1989, morgan2010}. In this case, a laminar $\Mach_\infty=2$ boundary layer over an adiabatic plate is separated by an impinging shock with a shock angle of $\theta \approx 32\text{\textdegree}$. The Reynolds number based on the impinging location of the shock is $\Reynolds_{x_0}=3\cdot10^5$. All other flow conditions match the simulation by \citet{katzer1989}. In the top-sponge opposite of the wall, Rankine-Hugoniot oblique shock relations are used to propagate the oblique shock downward, progressively introducing the discontinuity into the domain.

The grid used in that case is finer than in the previous studies, using respectively 960 and 501 points in the streamwise and wall-normal directions. All grid parameters are specified in \cref{table:mesh_swbli}.

\begin{figure}[tbph]
    \centering
    \includegraphics[width=0.99\textwidth]{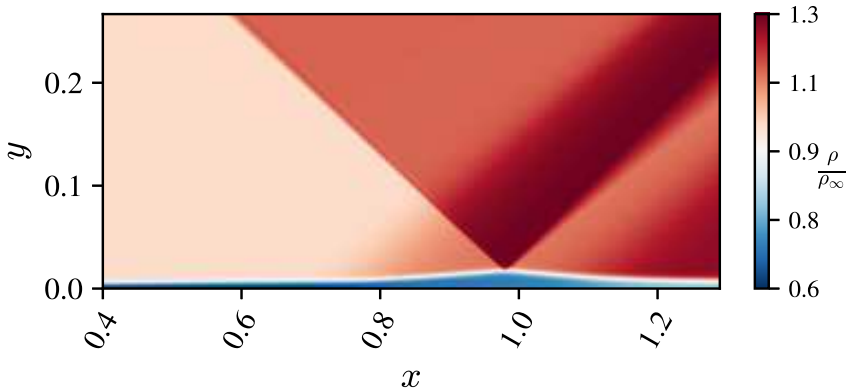}
    \caption{Density contour for the $\Mach_\infty=2$ laminar \SBLI\ case. The impinging and reflected shockwaves are visible, as well as the recirculation bubble and other relevant flow features.}
    \label{fig:rho_swbliM2}
\end{figure}

\begin{figure}[tbph]
    \centering
    \begin{subfigure}[t]{0.49\textwidth} 
        \centering \includegraphics[width=\textwidth]{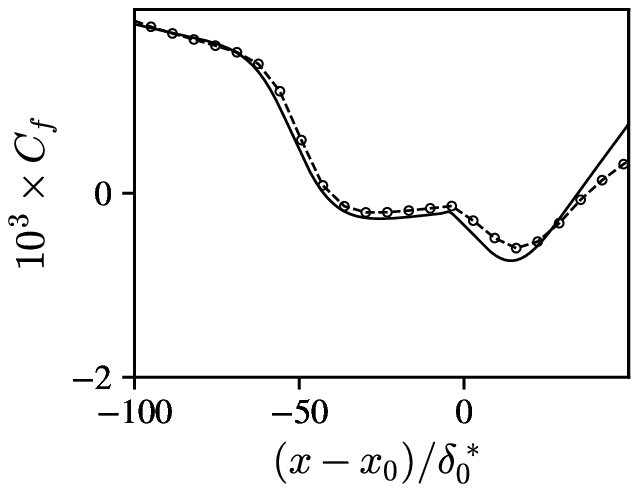}
        \caption{}\label{fig:cf_M2}
    \end{subfigure}
    \hfill
    \begin{subfigure}[t]{0.49\textwidth}
        \centering \includegraphics[width=\textwidth]{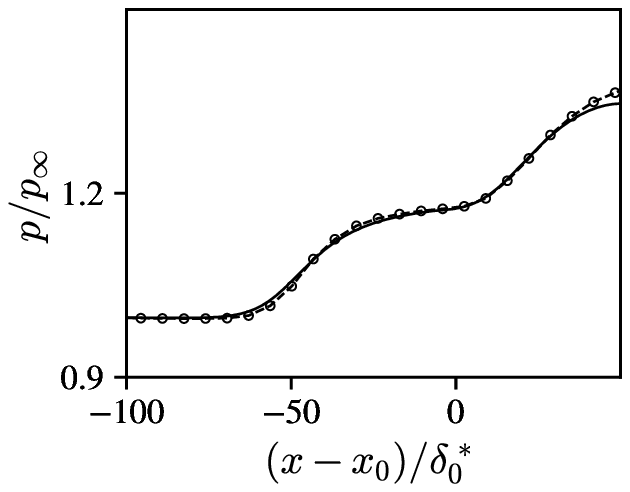}
        \caption{}\label{fig:pres_M2}
    \end{subfigure}
    \caption{Verification of the $\Mach_\infty=2$ laminar \SBLI\ case against literature. (a) Skin friction coefficient and (b) normalized wall pressure streamwise distributions. The impinging shock location is marked with $x_0$, and coincides with the location of zero on the $x$-axis. Current results (solid black line) are compared to \citeauthor{morgan2010}~\citep{morgan2010} (dashed lines with symbols).}\label{fig:cf_pres_swbliM2}
\end{figure}

\Cref{fig:rho_swbliM2} shows the density field after convergence of the residuals to machine-precision. All relevant flow features of the \SBLI\ are present: the recirculation bubble, the separation shock, the expansion fan, and the reflected shock. The results compare well to those of \cite{morgan2010}. 
\Cref{fig:cf_pres_swbliM2} presents the skin friction coefficient and wall pressure along the wall. These match previous results almost perfectly.

\subsubsection{2D \texorpdfstring{\SBLI\ at $\Mach=5.92$}{SBLI at Ma = 5.92}}
\Citet{furumoto1997} studied the real-gas effects on a steady oblique \SBLI\ at $\Mach=7$. The high enthalpy at the freestream leads to Oxygen and Nitrogen dissociation and reduction of the size of the recirculation region as well as peak heating on the surface due to the endothermic real-gas effects. However, \citeauthor{furumoto1997} note that the thermochemical model used was rather simplistic. 
More recently, \citet{volpiani2021} studied an oblique \SBLI\ at $\Mach=6$ in chemical non-equilibrium with both laminar and turbulent inflow boundary layers. These results highlight the same trend, with a smaller recirculation bubble and higher skin-friction at the wall when thermochemistry is considered in the model. However, the thermochemical model was again simplistic compared to the one included in \acrshort{Mutation++}.  \\
In this section, we propose to study the effects of finite-rate chemistry on a steady \SBLI\ at $\Mach=5.92$. A similar case was studied numerically in \cite{hildebrand2018} where the authors used freestream cryogenic conditions of the ACE Hypersonic Wind Tunnel facility at Texas A\&M University \cite{semper2012}. 
In this work, the freestream Mach number and Reynolds number at impinging location $\Reynolds_{x_0}$ were kept the same as in \citep{hildebrand2018}, while the freestream pressure and temperature have been increased to match the post-shock conditions of a 15\textdegree wedge flying at $\Mach=14$ at an altitude of 25 kilometers to promote real-gas effects.
The computational domain is a rectangle of size $256 \times 36 $ reference length units. The reference length is computed from the inflow Reynolds number $\Reynolds_{\delta^*}=9660$ of the original study \cite{hildebrand2018}, with the updated freestream conditions, $\freestream{\pressure}=\SI{60967.0}{Pa}$ and $\freestream{\temperature}= \SI{1110.5}{K}$, resulting in a similar impinging location and corresponding Reynolds number, $\Reynolds_{x_0}=1.15 \cdot 10^6$. A total of 960 grid points are used in the streamwise direction and 450 in the wall-normal direction, clustered near the wall using \cref{eq:stretching}. The mesh configuration is specified in \cref{table:mesh_swbli}. For this case, both the \TPG\ and \CNEQ\ models are investigated.

\begin{figure}[tbph]
    \centering
    \begin{subfigure}[t]{0.49\textwidth} 
        \centering \includegraphics[width=\textwidth]{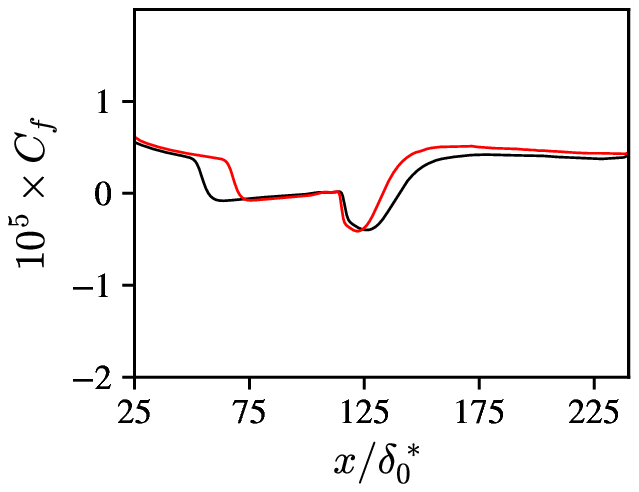}
        \caption{}\label{fig:cf_M6}
    \end{subfigure}
    \hfill
    \begin{subfigure}[t]{0.49\textwidth}
        \centering \includegraphics[width=\textwidth]{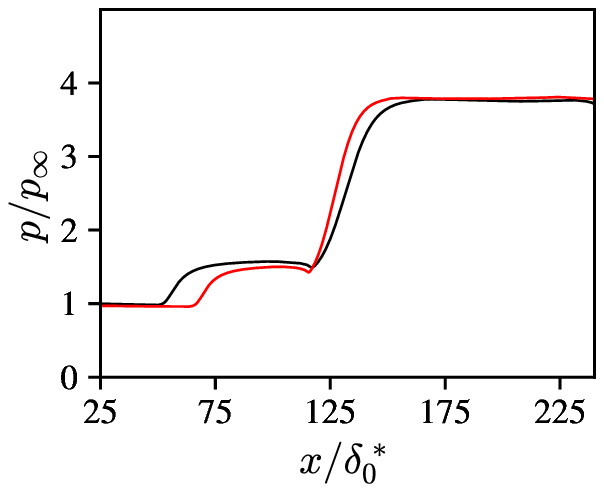}
        \caption{}\label{fig:pres_M6}
    \end{subfigure}
    \newline
    \begin{subfigure}[t]{0.49\textwidth}
        \centering \includegraphics[width=\textwidth]{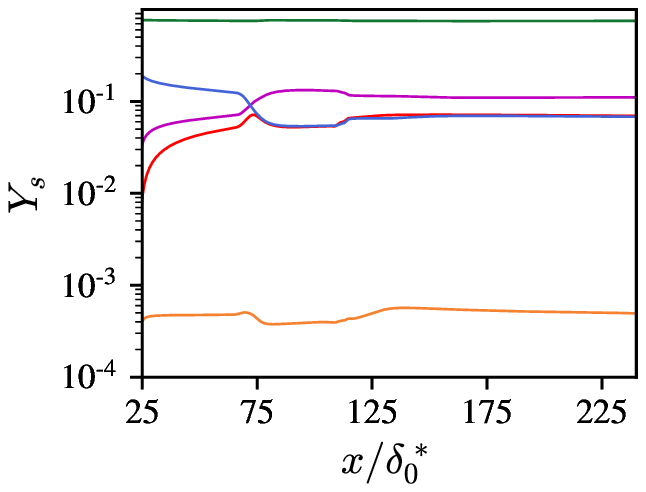}
        \caption{}\label{fig:species_M6}
    \end{subfigure}
    \caption{(a) Skin-friction coefficient and (b) normalized wall pressure streamwise distribution for the $\Mach_\infty=5.92$ \SBLI. Comparison of \TPG\ (black lines) and \CNEQ\ (red lines) results. (c) Mass fraction distributions at the wall for the \SBLI\ in \CNEQ. From top to bottom : $N_2$, $O_2$, $O$, $NO$, $N$.}\label{fig:cf_pres_swbliM6}
\end{figure}

The skin friction coefficient and wall pressure for both the \TPG\ and \CNEQ\ models are presented in \cref{fig:cf_pres_swbliM6}. The length of the separation bubble is smaller when considering \CNEQ\ effects, in agreement with literature. 
This is expected with the high concentration of dissociated species in the bubble as seen in \cref{fig:species_M6}. Just upstream of the reattachment location, the flow exhibits higher skin-friction and wall pressure in the \CNEQ\ case before converging to the same value as the \TPG\ case after reattachment. These trends are in agreement with the results in \cite{volpiani2021}.

\subsection{Jet in hypersonic crossflow}\label{sub:jetincrossflow}

Normal jet injection into a high-speed crossflow is another canonical flow configuration relevant to a wide range of applications in hypersonic flight. For example, proposed designs of scramjet engines use sonic under-expanded jet injection into a supersonic crossflow to enhance fuel and oxidizer mixing and sustain supersonic combustion. One can also use jet injection into a supersonic/hypersonic crossflow as a \RCS\ for aerodynamic maneuvering during atmospheric flight \cite{grandhi2017}. The jet in supersonic crossflow has also been the subject of a wide range of experimental \cite{erdem2011, santiago1997, ben2006} and numerical studies \cite{chai2015, kawai2010JISC, peterson2006}. The results presented in this section are two-dimensional. Thus, some similarity can be found with the interaction of hypersonic flow over an axisymmetric double cone \cite{tumuklu2018unsteadiness}.

The nondimensional parameter that governs the development of flow features is the jet to crossflow momentum ratio $J$ \cite{karagozian2014}, given as,
\begin{equation*}
J = \frac{\density_j U_j^2}{\density_\infty U_\infty^2}.
\end{equation*}

A case is designed by considering a $15\text{\textdegree}$ wedge flying at $\Mach=9$ at an altitude of 25 kilometers. This configuration results in a $\Mach=5$ boundary layer past the nose shock. A self-similar solution with freestream values assigned to post-shock conditions and an adiabatic wall results in $\temperature_\subs{wall} \approx 4000$ Kelvin (K). This temperature is sufficiently hot to observe intensified dissociation of $\mathrm{O}_{2}$ into atomic Oxygen and thus can result in a flow in \CNEQ. For this case, the \CNEQ\ and \TPG\ models are considered.
The freestream and jet thermodynamic conditions are summarized in \cref{table:condition_JISC}.
The computational domain is a rectangle of size $32.5D \times 10D $ where $\tilde{D} = \SI{2}{mm}$ is the jet slot width. In the streamwise direction, the inflow starts at $x_0 = 38.5 $ and the injection zone is around $x = 52.5$.  The inflow and outflow sponges extend for 1 and 2 jet widths, respectively. In the wall-normal direction, grid points are clustered near the wall using \cref{eq:stretching}, and the last 20 points are used for the sponge. The discretization for this case is specified in \cref{table:mesh_jisc}. The simulation is first initialized with an adiabatic condition and the resulting wall temperature is held constant after the jet injection is enforced.

\begin{table}[tbph]
\begin{center}
\begin{tabular}{ |c|c|c|c||c|c|c|c| } 
 \hline
 \multicolumn{4}{|c||}{Freestream} & \multicolumn{4}{c|}{Jet} \\
 \hline
 $\Mach_\freest$ & $\freestream{\pressure} [Pa]$ & $\freestream{\temperature} [K]$ & $\Reynolds_{D}$ & $\Mach_\subs{jet}$ &  $\tilde{\pressure}_\subs{jet}/\freestream{\pressure}$ & $\tilde{\temperature}_\subs{jet}/\freestream{\temperature} $ & $J$\\ 
 \hline
 5 & 49800 & 947 & 25291 & 1.0 & 10.0 & 1.0 & 0.4  \\ 
 \hline
\end{tabular}
\caption{Thermodynamic conditions for the \acrshort{2D} jet in hypersonic crossflow.}
\label{table:condition_JISC}
\end{center}
\end{table} 
 
\begin{table}[tbph]
\begin{center}
\begin{tabular}{ |l|| c c | c c |c c| } 
 \hline
 Case & $x_0 - x_1$ & $y_0 - y_1$ & $\kappa_x$ & $\kappa_y$ & $N_x$ & $N_y$ \\ 
 \hline
 JISC & $38.5 - 71.0$ & $0.0 - 10.0$ & $ 1.0 $ & $ 0.15 $ & 1625 & 500 \\
  \hline
\end{tabular}
\caption{Mesh configuration for the jet in hypersonic crossflow simulation.}
\label{table:mesh_jisc}
\end{center}
\end{table}

\begin{figure}[hbtp]
    \centering
    \begin{subfigure}[t]{0.99\textwidth} 
        \centering \includegraphics[width=\textwidth]{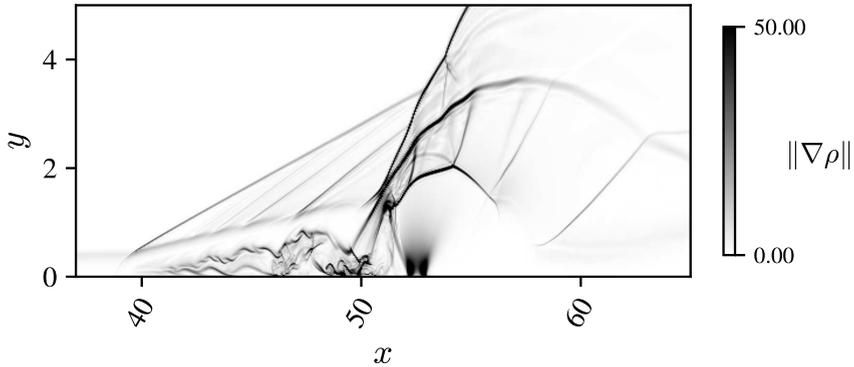}
        \caption{}\label{fig:schlieren_cold}
    \end{subfigure}
    \begin{subfigure}[t]{0.99\textwidth}
        \centering \includegraphics[width=\textwidth]{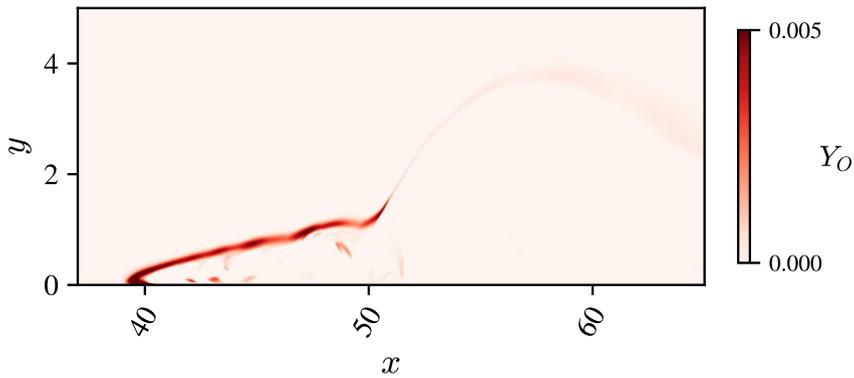}
    \caption{} \label{fig:chemical_JISC}
    \end{subfigure}
    
    \caption{Instantaneous (a) numerical Schlieren and (b) atomic Oxygen mass fraction fields for the sonic jet in $\Mach_\infty=5$ hypersonic crossflow. Jet location is at $x=52.5$. Configuration parameters may be found in \cref{table:condition_JISC} and \cref{table:mesh_jisc}.} \label{fig:Cold_JISC_mach_vort}
\end{figure}

\Cref{fig:schlieren_cold} shows a numerical Schlieren image (contours of the density gradient norm) for the jet in crossflow simulation in \CNEQ. A number of characteristic flow features are observed. The incoming laminar boundary layer encounters separation upstream of the injection zone, induced by the separation shock created by the upstream unsteady recirculation bubble. A large bow-shock forms in the freestream that diverts the flow. Moreover, a shear layer emanates from the jet shock structures and the flow past the bow shock induces vortex shedding. This results in a strong coupling between the shock structures, the recirculation bubble, and the shear layer downstream of the injection zone. 
The structure observed and these interactions are thus similar to those observed for a $\Mach = 16$ axisymmetric shock-dominated hypersonic laminar separated flow over a double cone studied by \citet{tumuklu2018unsteadiness}. In their study, a supersonic underexpanded jet is generated through an Edney type IV pattern in the shock-laminar separation bubble. The jet is inherently unsteady and becomes the root of the \SBLI\ instability.

\begin{figure}
    \centering
    \includegraphics[width=0.5\textwidth]{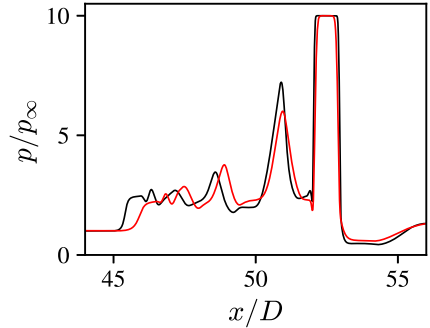}
    \caption{Instantaneous streamwise distribution of wall pressure for the sonic jet in $\Mach_\infty = 5$ crossflow simulation. Comparison of \TPG\ (black lines) and \CNEQ\ (red lines) results.}
    \label{fig:pres_JISC}
\end{figure}

Areas of high concentration of $\mathrm{O}$ are observed inside the recirculation bubble in \cref{fig:chemical_JISC}. In this high-temperature region, atomic Oxygen is produced through the endothermic $\mathrm{O}_2$ dissociation. This reaction absorbs energy from the flow and reduces the size of the recirculation bubble by 10\% compared to the \TPG\ simulation \cref{fig:pres_JISC}. This observation holds for the duration of the instantaneous snapshots considered and is similar to the trend observed in the steady $\Mach=5.92$ \SBLI\ comparison between the \TPG\ and \CNEQ\ models. When \CNEQ\ effects are considered, the adiabatic wall temperature decreases due to endothermic reactions (by about \SI{250}{K} at the wall) and the boundary layer becomes slightly thicker. The induced cooling effect near the wall leads to a smaller recirculation bubble, and a weaker bow shock. Future work will further analyze the effects of the finite-rate chemistry on the flow field compared to the \TPG\ thermodynamic model, as well as potential changes in the instability characteristics of the flow field in a more realistic 3D configuration.

\section{Conclusions}\label{sec:conclusions}

The major challenges posed by the need for space exploration and faster and higher air travel have undoubtedly reinvigorated research interest in applications of hypersonic flow. At such high speed and high enthalpy regimes such as the those encountered during atmospheric (re-)entry or sustained hypersonic cruise, gases do not behave in ways that are predicted by the typical models of \CFD\ solvers, and improved models are necessary. In addition, the complex flow scenario results in a large number of phenomena that interact in often non-trivial manners.

In this work, we have presented a computational tool that can accurately simulate flows at hypersonic speed, including high-enthalpy gas effects in the presence of weak and strong shockwaves. We summarize some important milestones in that direction toward our research objective, having verified our numerical tools and tested their application to more complex and physically relevant configurations. The agreement of the presented results is excellent, in particular in cases where chemistry and shockwaves significantly interact with the flow and physics.

Extensive code verification has been completed, against previous results for reacting boundary layers, \SBLI\ cases, and complex jet injections in high-speed crossflow. The applications have been extended in novel, higher-enthalpy cases where the potential effects of gas thermochemistry are pronounced.

There are a number of directions in which there is significant potential for improvement and extension of the tools’ capabilities. First, there are even more accurate options for gas modeling, including state-to-state transitions and taking into account chemistry at the wall (catalicity). Second, there is great interest in extending our results for different gas mixtures, representing atmospheric compositions of celestial objects interesting for space exploration, or for more complex geometry configurations. Third, on the algorithmic side, there are methodologies that can complement, or in some cases even replace, the computationally expensive tasks of the thermochemical model library, using tools from the fields of machine-learning and reduced-order modeling.

\backmatter

\bmhead{Acknowledgments}

The authors wish to acknowledge the support of the US Air Force Office of Scientific Research (AFOSR) and the European Office of Aerospace Research and Development (EOARD) under the grant FA9550-18-1-0127, managed by Dr. Sarah Popkin, Dr. Ivett Leyva, and Dr. Douglas Smith.
Computational time on National UK Supercomputers has been provided by the UK Turbulence Consortium (UKTC) and Dr. Sylvain Laizet, under the grant EP/R029326/1. Additional computational resources have been made available by the corresponding universities and agencies (Imperial College London, King Abdullah University of Science and Technology (KAUST), Sorbonne University, and the French National Research Center (CNRS)).

\bmhead{Competing interests}

The authors report no competing interests. This work was supported by AFOSR and EOARD (Grant FA9550-18-1-0127). Computational time was provided by UKTC (Grant EP/R029326/1). The views and conclusions presented herein are those of the authors and do not represent the views of any of the aforementioned organizations.

\bmhead{Code availability}

The \acrlong{Mutation++} library repository is publicly available on \url{https://github.com/mutationpp/Mutationpp}. The rest of the computational tools developed may be shared by the corresponding author on reasonable request.

\bmhead{Data availability}

The datasets generated and/or analyzed during the current study are available from the corresponding author on reasonable request.

\bibliography{references.bib}

\end{document}

%% file: preamble.tex
\usepackage[utf8]{inputenc}
\usepackage{graphicx}
\usepackage{caption}
\usepackage{subcaption}
\usepackage{amsmath}
\usepackage[version=4]{mhchem}
\usepackage{siunitx}
\usepackage{longtable,tabularx}
\usepackage{upgreek}
\usepackage{siunitx}
\usepackage{mathtools}
\usepackage{xcolor}
\usepackage{tikz}

\usepackage[capitalise,nameinlink,sort]{cleveref}
\crefrangeformat{equation}{#3Eqs.~(#1--#2)#6}
\crefname{paragraph}{Section}{Sections}
\Crefname{paragraph}{Section}{Sections}

\input{symbols}

\input{nomenclature}

\glsdisablehyper
\hypersetup{colorlinks=false,hidelinks}

%% file: symbols.tex
\usepackage{bm}

\newcommand{\pd}[2]{\ensuremath{\frac{\partial #1}{\partial #2}}}

\renewcommand{\vec}[1]{\ensuremath{\mathbf{#1}}}
\newcommand{\newvec}[1]{\ensuremath{\mathbf{#1}}}
\newcommand{\tens}[1]{\ensuremath{\mathrm{#1}}}

\newcommand{\subs}[1]{\ensuremath{\mathrm{#1}}}

\newcommand{\grad}[1]{\ensuremath{\vec{\nabla}#1}}

\newcommand{\divg}[1]{\ensuremath{\vec{\nabla}\cdot#1}}
\newcommand{\divgp}[1]{\ensuremath{\vec{\nabla}\cdot \left( #1 \right)}}

\newcommand{\norm}[1]{\ensuremath{\lvert\lvert #1 \lvert\lvert}}

\newcommand{\avg}[1]{\ensuremath{\bar{#1}}}
\newcommand{\lin}[1]{\ensuremath{#1}}

\newcommand{\transposep}[1]{\ensuremath{\left({#1}\right)^\mathrm{T}}}

\newcommand{\dimval}[1]{\ensuremath{\tilde{#1}}}
\newcommand{\freestream}[1]{\ensuremath{\dimval{#1}_\freest}}
\newcommand{\valref}[1]{\ensuremath{\dimval{#1}_\subsref}}

\newcommand{\mathset}[1]{\ensuremath{\mathcal{#1}}}

\newcommand{\sumspecies}{\sum_{\speciess\in\speciesset}}
\newcommand{\sumreactions}{\sum_{\reactionr\in\reactionsset}}
\newcommand{\eyematrix}{\tens{I}}

%% file: nomenclature.tex
\usepackage[section=section,nogroupskip,nonumberlist,acronym,nopostdot,hyperfirst=true]{glossaries}
\newglossarystyle{longRaggedRight2cm}{%
  \setglossarystyle{long}%
  \renewenvironment{theglossary}%
     {\begin{longtable}{p{2.5cm}p{\textwidth minus 3cm}}}%
     {\end{longtable}}%
}
\newglossarystyle{longRaggedRight1cm}{%
  \setglossarystyle{long}%
     {\begin{longtable}{p{1cm}p{\textwidth minus 1.5cm}}}%
     {\end{longtable}}%
}

\newglossary[clg]{const}{cns}{con}{Constants}
\newglossary[rlg]{roman}{rom}{rtn}{Roman symbols}
\newglossary[slg]{subsc}{sub}{stn}{Subscripts}
\newglossary[glg]{greek}{grk}{grn}{Greek symbols}

\makeglossaries

\newcommand{\linkvartoglossary}[4]{
	\newglossaryentry{#2}{type=#1,name={\ensuremath{#3}},description={#4},sort={#3}}%
    \expandafter\newcommand\csname #2\endcsname{{\gls{#2}}}%
    \expandafter\newcommand\csname #2i\endcsname[1]{{\gls{#2}[_##1]}}%
    \expandafter\newcommand\csname #2s\endcsname[1]{{\gls{#2}[^##1]}}%
}

\newcommand{\addconst}[3]{\linkvartoglossary{const}{#1}{#2}{#3}}
\newcommand{\addroman}[3]{\linkvartoglossary{roman}{#1}{#2}{#3}}
\newcommand{\addsubsc}[3]{\linkvartoglossary{subsc}{#1}{#2}{#3}}
\newcommand{\addgreek}[3]{\linkvartoglossary{greek}{#1}{#2}{#3}}
\newcommand{\addacronym}[2]{
	\newacronym{#1}{#1}{#2}%
	\expandafter\newcommand\csname #1\endcsname{{\gls{#1}}}%
}
\newcommand{\adddualacronym}[3]{
	\newacronym[description=#3]{#1}{#1}{#2}%
	\expandafter\newcommand\csname #1\endcsname{{\gls{#1}}}%
}

\addconst{RU}{\mathcal{R}_\subs{u}}{}

\addsubsc{speciess}{\subs{s}}{quantity refers to species $\mathrm{s}$}
\addsubsc{directioni}{\subs{i}}{quantity refers to direction $\mathrm{i}$}
\addsubsc{totalnot}{\subs{0}}{quantity refers to total or stagnation (static plus dynamic)}
\addsubsc{freest}{\subs{\infty}}{quantity refers to the free stream at the inlet}
\addsubsc{subsref}{\subs{ref}}{quantity refers to the free stream at the inlet}

\addroman{soundspeed}{c}{speed of sound}
\addroman{pressure}{p}{pressure}
\addroman{temperature}{T}{temperature}
\addroman{mw}{M}{}
\addroman{mwspecies}{\mw_\speciess}{}

\addroman{numspecies}{N_\speciess}{number of species}

\addroman{Eckert}{Ec}{Eckert number at free stream, at inlet, \Eckert = }
\addroman{Mach}{Ma}{Mach number at free stream, at inlet, \Mach = }
\addroman{Reynolds}{Re}{Reynolds number at free stream, at inlet, 
$\Reynolds = \freestream{\density} \freestream{\soundspeed} \valuereference{L} / \valuereference{\viscosity}$}
\addroman{Prandtl}{Pr}{Prandtl number at free stream, at inlet, \Prandtl = }

\addroman{viscosity}{\mu}{viscosity}
\addroman{thermconductivity}{k}{thermal conductivity}
\addroman{heatp}{c_p}{thermal conductivity}

\addroman{internale}{e}{internal specific energy}
\addroman{internalh}{h}{internal specific enthalpy}
\addroman{totale}{\internale_\totalnot}{total specific energy (internal and kinetic)}
\addroman{totalh}{\internalh_\totalnot}{total specific enthalpy (internal and kinetic)}

\addroman{speciese}{\internale_\speciess}{internal specific energy of species $\speciess$}
\addroman{speciesh}{\internalh_\speciess}{internal specific enthalpy of species $\speciess$}

\addroman{speciesmassfrac}{Y_\speciess}{mass fraction of species $\speciess$}
\addroman{speciesmolefrac}{X_\speciess}{mole fraction of species $\speciess$}

\addroman{timevar}{t}{time variable}
\addroman{velocity}{\vec{u}}{hydrodynamic velocity (vector)}
\addroman{veli}{u_i}{hydrodynamic velocity in direction $\directioni$ (scalar)}
\addroman{diffvelocity}{\vec{V}_\speciess}{diffusion velocity for species $\speciess$}

\addroman{heatflux}{\vec{q}}{heat flux vector}

\addroman{statevec}{\newvec{Q}}{}
\addroman{statevecavg}{\avg{\newvec{Q}}}{}
\addroman{stateveclin}{\lin{\newvec{q}}}{}
\addroman{stateveclinhat}{\hat{\stateveclin}}{}

\addroman{fluxvec}{\vec{F}}{}
\addroman{fluxvecconv}{\vec{F}_\subs{C}}{}
\addroman{fluxvecdiff}{\vec{F}_\subs{D}}{}

\addroman{sourcevec}{\vec{S}}{}

\addroman{jacobianA}{\tens{A}}{}
\addroman{jacobianAhat}{\hat{\jacobianA}}{}
\addroman{hessH}{\tens{H}}{}
\addroman{hessHhat}{\hat{\tens{H}}}{}
\addroman{jacobian}{\pd{\fluxvec}{\statevec}}{}

\addroman{resolventR}{\tens{R}_\subs{\jacobianA}}{}
\addroman{svddiag}{\tens{\Sigma}}{}
\addroman{svdin}{\tens{V}}{}
\addroman{svdout}{\tens{U}}{}

\addroman{speciesset}{\mathset{S}}{}
\addroman{reactionsset}{\mathset{R}}{}

\addroman{reactionr}{\subs{r}}{}
\addroman{reactionrate}{R_\reactionr}{}
\addroman{forwardrate}{{k_f}_\reactionr}{}
\addroman{backwardrate}{{k_b}_\reactionr}{}
\addroman{equilconst}{{K_\subs{eq}}_\reactionr}{}

\addroman{forcing}{\vec{f}}{}

\addroman{spacevec}{\vec{x}}{}

\addgreek{massprod}{{\dot{\omega}}_\speciess}{mass production rate due to chemical reactions}
\addgreek{density}{\rho}{density}
\addgreek{densityspecies}{\density_\speciess}{density of species $\speciess$}
\addgreek{viscstresstensor}{\tens{\uptau}}{viscous stress tensor}

\addgreek{gammaratio}{\gamma}{specific heat ratio}

\addgreek{sponge}{\sigma}{}

\adddualacronym{2D}{two-dimensional}{Two-dimensional}
\adddualacronym{3D}{three-dimensional}{Three-dimensional}

\addacronym{AIAA}{American Institute of Aeronautics and Astronautics}

\addacronym{AFOSR}{Air Force Office for Scientific Research}
\addacronym{EOARD}{European Office of Aerospace Research and Development}

\adddualacronym{CFD}{computational fluid dynamics}{Computational Fluid Dynamics}
\addacronym{CFL}{Courant-Friedrichs-Lewy}
\adddualacronym{CNEQ}{chemical non-equilibrium}{Chemical Non-Equilibrium}
\adddualacronym{CPG}{calorically perfect gas}{Calorically Perfect Gas (often called ideal or perfect gas)}

\adddualacronym{DMD}{dynamic mode decomposition}{Dynamic Mode Decomposition}
\adddualacronym{DNS}{direct numerical simulation}{Direct Numerical Simulation}
\adddualacronym{DPLR}{data-parallel line relaxation}{Data-Parallel Line Relaxation (numerical method or computational code)}

\adddualacronym{IBM}{immersed boundary method}{Immersed Boundary Method}

\adddualacronym{LES}{large-eddy simulation}{Large-Eddy Simulation}
\adddualacronym{LST}{linear stability theory}{Linear Stability Theory}
\adddualacronym{LTE}{local thermodynamic equilibrium}{Local Thermodynamic Equilibrium}

\adddualacronym{MFP}{mean-flow perturbation}{Mean-Flow Perturbation}

\adddualacronym{MT}{multi-temperature}{Multi-Temperature}
\adddualacronym{MUTATION}{MUTATION}{MUlticomponent Thermodynamic And Transport properties for IONized gases (library)}
\adddualacronym{Mutation++}{Mutation++}{Multicomponent thermodynamic and transport properties for ionized gases in C++ (library)}

\addacronym{NASA}{National Aeronautics and Space Administration}
\adddualacronym{NEQ}{non-equilibrium}{Non-Equilibrium}

\adddualacronym{PSE}{parabolized stability equations}{Parabolized Stability Equations}

\adddualacronym{RCS}{reaction control system}{Reaction Control System}

\adddualacronym{SBLI}{shockwave--boundary-layer interaction}{Shockwave--Boundary-Layer Interaction}
\adddualacronym{SVD}{singular value decomposition}{Singular Value Decomposition}

\adddualacronym{TCNEQ}{thermal and chemical non-equilibrium}{Thermal and Chemical Non-Equilibrium}
\adddualacronym{TNEQ}{thermal non-equilibrium}{Thermal Non-Equilibrium}
\adddualacronym{TPG}{thermally perfect gas}{Thermally Perfect Gas (often called imperfect gas)}
\adddualacronym{TPS}{thermal protection system}{Thermal Protection System}
\adddualacronym{TVD}{total variation diminishing}{Total Variation Diminishing}

\adddualacronym{US3D}{{US3D}}{{US3D} code{,} developed by University of Minnesota and NASA}

\adddualacronym{VKI}{von Karman Institute for Fluid Dynamics}{von Karman Institute for Fluid Dynamics (Belgium)}

%% file: Figures/Fig1.tex
\definecolor{myred}{RGB}{180,50,50}
\definecolor{myblue}{RGB}{50,50,180}
\definecolor{mygreen}{RGB}{90,160,20}
\definecolor{mygray}{RGB}{120,120,120}

\begin{tikzpicture}[scale=1]

    \draw[step=1cm, mygray, thin, shift={(0.0,0.0)}] (0,0) grid (3.7,3.7);

    \def\nodelist{0,...,3}
    \def\alength{0.25}

    \draw [->, >=latex, mygray, thick] (0,0) -- (4,0) node[anchor=west] {$x$};
    \draw [->, >=latex, mygray, thick] (0,0) -- (0,4) node[anchor=south] {$y$};
    
    \foreach \i in \nodelist{
        \foreach \j in \nodelist{
            \node[fill, black, inner sep=1.5pt] at (\i,\j) {};
            \node[circle, fill, myblue, inner sep=1.2pt] at (\i+0.5,\j+0.5) {};
            \draw [->, >=latex, thick, myred]   (\i-\alength/2,\j+0.5) -- (\i+\alength/2,\j+0.5) {};
            \draw [->, >=latex, thick, mygreen] (\i+0.5,\j-\alength/2) -- (\i+0.5,\j+\alength/2) {};
        }
    }

    \node[fill, black, inner sep=1.5pt, label={[label distance=0cm]0:\makebox[1.4cm][r]{$\tau$ node}}] at (4.5,3) {};
    \node[fill, myblue, circle, inner sep=1.2pt, label={[label distance=0cm]0:\makebox[1.4cm][r]{\color{myblue}$\rho$ node}}] at (4.5,1.5) {};
    \draw [->, >=latex, thick, myred]   (4.5-\alength/2,2.5) -- (4.5+\alength/2,2.5) node[midway, align=right, rotate=0, anchor=west] 
    {\makebox[1.4cm+1.5pt][r]{$\;\rho\vec{u}$ node}};
    \draw [->, >=latex, thick, mygreen] (4.5,2-\alength/2) -- (4.5,2+\alength/2) node[midway, align=right, rotate=0, anchor=west]
    {\makebox[1.4cm+1.5pt][r]{$\;\rho\vec{v}$ node}};

\end{tikzpicture}

%% file: Figures/Fig2.tex
\definecolor{myred}{RGB}{180,50,50}
\definecolor{myblue}{RGB}{50,50,180}
\definecolor{mygreen}{RGB}{90,160,20}
\definecolor{mygray}{RGB}{120,120,120}

\begin{tikzpicture}[scale=1]

    \def\alength{0.15}
    \def\astep{0.5}
    
    \def\spongethickness{0.5}
    \def\domainwidth{10.0}
    \def\domainheight{2.0}
    
    \fill[mygray, opacity=0.25] (0,0) rectangle (\spongethickness,\domainheight);
    \fill[mygray, opacity=0.25] (\domainwidth-\spongethickness,0) rectangle (\domainwidth,\domainheight);
    \fill[mygray, opacity=0.25] (\spongethickness,\domainheight-\spongethickness) rectangle (\domainwidth-\spongethickness,\domainheight);
    \node[anchor=center, fill=white, inner sep=1pt, fill opacity=0.95] at (\domainwidth/2.0,\domainheight-0.5*\spongethickness) {\color{mygray} Sponge regions};
    \draw[thin, dashed, black] (\spongethickness,0) -- (\spongethickness,\domainheight-\spongethickness) -- (\domainwidth-\spongethickness,\domainheight-\spongethickness) -- (\domainwidth-\spongethickness,0);

    \draw [->, >=latex, mygray, thick] (0,0) -- (\domainwidth+\spongethickness,0) node[anchor=west]  {$x$};
    \draw [->, >=latex, mygray, thick] (0,0) -- (0,\domainheight+\spongethickness) node[anchor=south] {$y$};
    
    \draw[very thick, black]   (0,0) -- (\domainwidth,0) {}; 
    \foreach \i in {0,\astep,...,\domainwidth}{ \draw[thick, black] (\i,0) -- (\i+1.5*\alength,-\alength) {}; }
    \node[anchor=north, fill=white, inner sep=1pt, fill opacity=0.95] at (\domainwidth/2.0,-\spongethickness/2.0) {\color{black} Wall}; 
    \draw[very thick, myred]   (0,0) -- (0,\domainheight) {}; \node[anchor=east]  at (0, \domainheight/2.0) {\color{myred}   Inflow}; 
    \draw[very thick, myblue]  (0,\domainheight) -- (\domainwidth,\domainheight) {}; \node[anchor=south] at (\domainwidth/2.0,\domainheight) {\color{myblue}  Unperturbed free-stream}; 
    \draw[very thick, mygreen] (\domainwidth,0) -- (\domainwidth,\domainheight) {}; \node[anchor=west]  at (\domainwidth,\domainheight/2.0) {\color{mygreen} Outflow};

    \draw[black] (0,\domainheight/10.0) .. controls (\domainwidth/10.0,\domainheight/7.0) and (\domainwidth/5.0,\domainheight/5.0) .. (\domainwidth,\domainheight/4.0) {};
    
    \draw[->, >=latex, black] (\domainwidth/2.5,\domainheight/2.0) .. controls (\domainwidth/2.0,\domainheight/2.0) and (\domainwidth/1.5,\domainheight/2.0) .. (\domainwidth/1.5,\domainheight/3.5) node[pos=0, anchor=east] {Boundary-layer edge};

\end{tikzpicture}